\RequirePackage{fix-cm}
\documentclass[twocolumn]{svjour3}          
\usepackage{algorithm}
\usepackage{algorithmic}
\smartqed  
\usepackage{ragged2e}
\usepackage{booktabs,makecell,multirow,tabularx}
\usepackage{amsmath}
\usepackage{graphicx}
\usepackage{amssymb}
\usepackage{xcolor}
\usepackage{colortbl,booktabs}
\begin{document}
\begin{sloppypar}
\title{DmADs-Net: Dense multiscale attention and depth-supervised network for medical image segmentation
}
\subtitle{}


\author{Zhaojin Fu
	  \and
	  Zheng Chen
	  \and
      Jinjiang Li
      \and
      Lu Ren
}


\institute{
	Zhaojin Fu is with School of Computer Science and Technology, Shandong Technology and Business University, Yantai 264005, China.
	\email{1226187850@qq.com}
	\and
	\\
	Zheng Chen is with School of Computer Science and Technology, Shandong Technology and Business University, Yantai 264005, China.
	\email{chenzheng@sdtbu.edu.cn}
	\\
    Jinjiang LI is with School of Computer Science and
	Technology, Shandong Technology and Business University, Yantai 264005, China.
	\email{lijinjiang@gmail.com}
	\and
	\\
	Lu Ren is with School of Computer Science and Technology, Shandong Technology and Business University, Yantai 264005, China.
	\email{renlu@sdtbu.edu.cn}
}
\date{Received: date / Accepted: date}

\maketitle

\begin{abstract}
Deep learning has made important contributions to the development of medical image segmentation. Convolutional neural networks, as a crucial branch, have attracted strong attention from researchers. Through the tireless efforts of numerous researchers, convolutional neural networks have yielded numerous outstanding algorithms for processing medical images. The ideas and architectures of these algorithms have also provided important inspiration for the development of later technologies.Through extensive experimentation, we have found that currently mainstream deep learning algorithms are not always able to achieve ideal results when processing complex datasets and different types of datasets. These networks still have room for improvement in lesion localization and feature extraction. Therefore, we have created the Dense Multiscale Attention and Depth-Supervised Network (DmADs-Net).We use ResNet for feature extraction at different depths and create a Multi-scale Convolutional Feature Attention Block to improve the network's attention to weak feature information. The Local Feature Attention Block is created to enable enhanced local feature attention for high-level semantic information. In addition, in the feature fusion phase, a Feature Refinement and Fusion Block is created to enhance the fusion of different semantic information.We validated the performance of the network using five datasets of varying sizes and types. Results from comparative experiments show that DmADs-Net outperformed mainstream networks. Ablation experiments further demonstrated the effectiveness of the created modules and the rationality of the network architecture.
\end{abstract}

\begin{figure*}[!t]
	\centering
	\includegraphics[width=\linewidth]{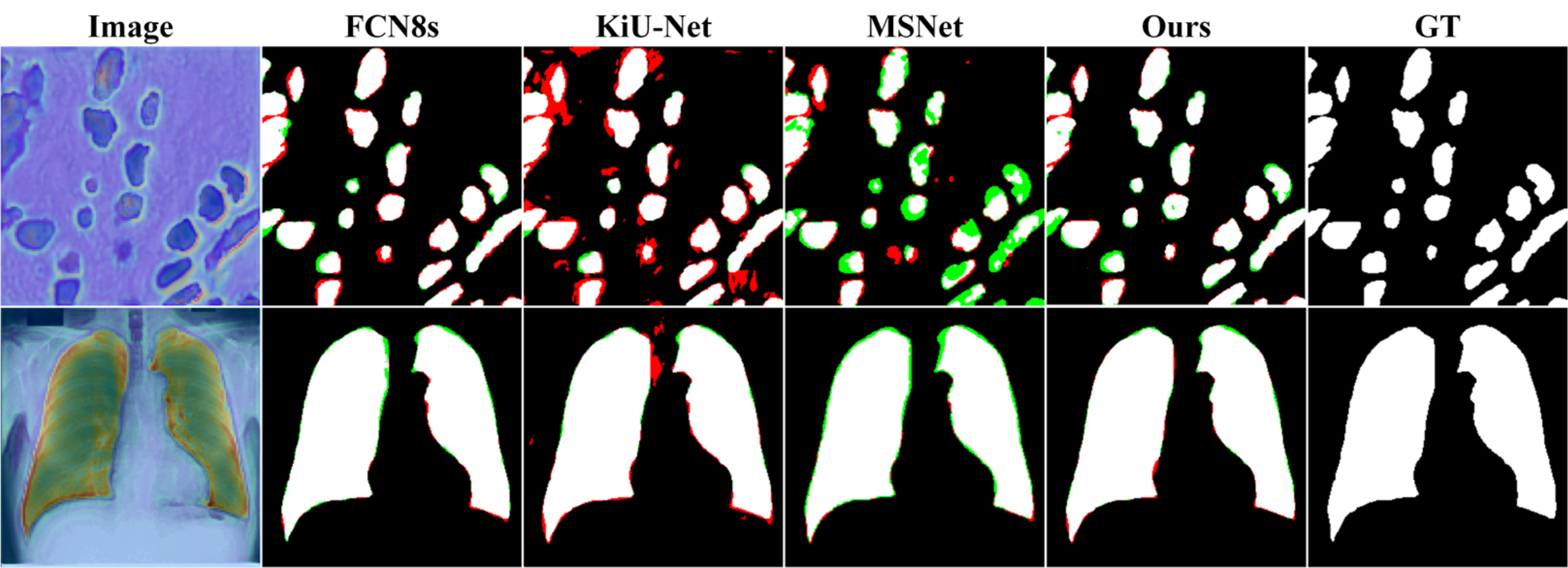}
	\caption{A comparative experiment diagram of Ours and mainstream networks.We annotate the attention features of the segmentation targets in the images and perform visual overlap processing for all participating methods.Red for incorrectly identified areas, green for unidentified areas.}
	\label{fig1}
\end{figure*}
\section{Introduction}
The analysis of a lesion requires the doctor to determine the shape, size and contour of the lesion. Lesions on the outside of the body can be identified by direct observation, but lesions on the inside of the body can only be analysed indirectly by means of images transmitted from the imaging equipment. Although the above two processes complete the analysis of the lesion, they consume a lot of time and effort. Therefore, autonomous learning for image segmentation tasks has become an important research goal in the medical field.

In computer-aided diagnosis (CAD) systems, automatic image segmentation algorithms provide important information and data for the entire diagnostic process as an upstream task. Particularly in quantitative and qualitative analysis, high-quality segmentation data helps doctors to make efficient and quick diagnoses.

Most of the early medical image segmentation tasks are done using traditional algorithms\cite{1,2,3}, where different regions of a medical image are divided by means of pixel changes and threshold processing to localise the target of a lesion. However, with the advancement of the medical field, the clarity and precision of medical images have continuously improved, making the segmentation tasks faced by traditional algorithms more diversified and complex. Although traditional algorithms have made important contributions to medical image segmentation tasks, relying solely on thresholding and pixel variations is no longer sufficient to meet the diagnostic needs of medical care. Deep learning provides new solutions for medical image segmentation tasks. More and more research workers began to pay attention to the development of deep learning\cite{56,57,58,59} in the field of medical image segmentation has given birth to many excellent algorithms\cite{4,5,6}, in which convolutional neural network (CNN) is the most representative of a branch.

Convolutional neural networks such as AlexNet\cite{7} process the features generated by convolution into feature vectors for performing classification tasks through fully connected layers. However, this all-connected layer approach does not apply in the case of segmentation tasks. The emergence of FCN\cite{8} has brought the field of image segmentation into a new period. In order to be able to achieve pixel-level classification and solve image segmentation problems, FCNs changed the way the network ended up using fully connected layers and used convolution instead. This change provides an important lesson in thinking for the field of medical image segmentation. Subsequently, SegNet\cite{9} explored semantic segmentation based on the network of VGG\cite{10} and proposed the "encoder-decoder" structure for the first time. The size is reduced by maximum pooling during the encoding phase and feature information at different semantic levels is continuously acquired through convolutional layers. The feature image is restored to the same size in the decoding stage by upsampling to the same position as in the encoding stage. Finally, pixel-level classification of the feature maps of the original size is performed, which enables the segmentation of the whole image. With the development of technology and ideas, networks such as U-Net\cite{11}, UNet++\cite{12} and KiU-Net\cite{13} have emerged one after another, continuously pushing forward the quality of medical image segmentation.

\begin{figure*}[!t]
	\centering
	\includegraphics[width=\linewidth]{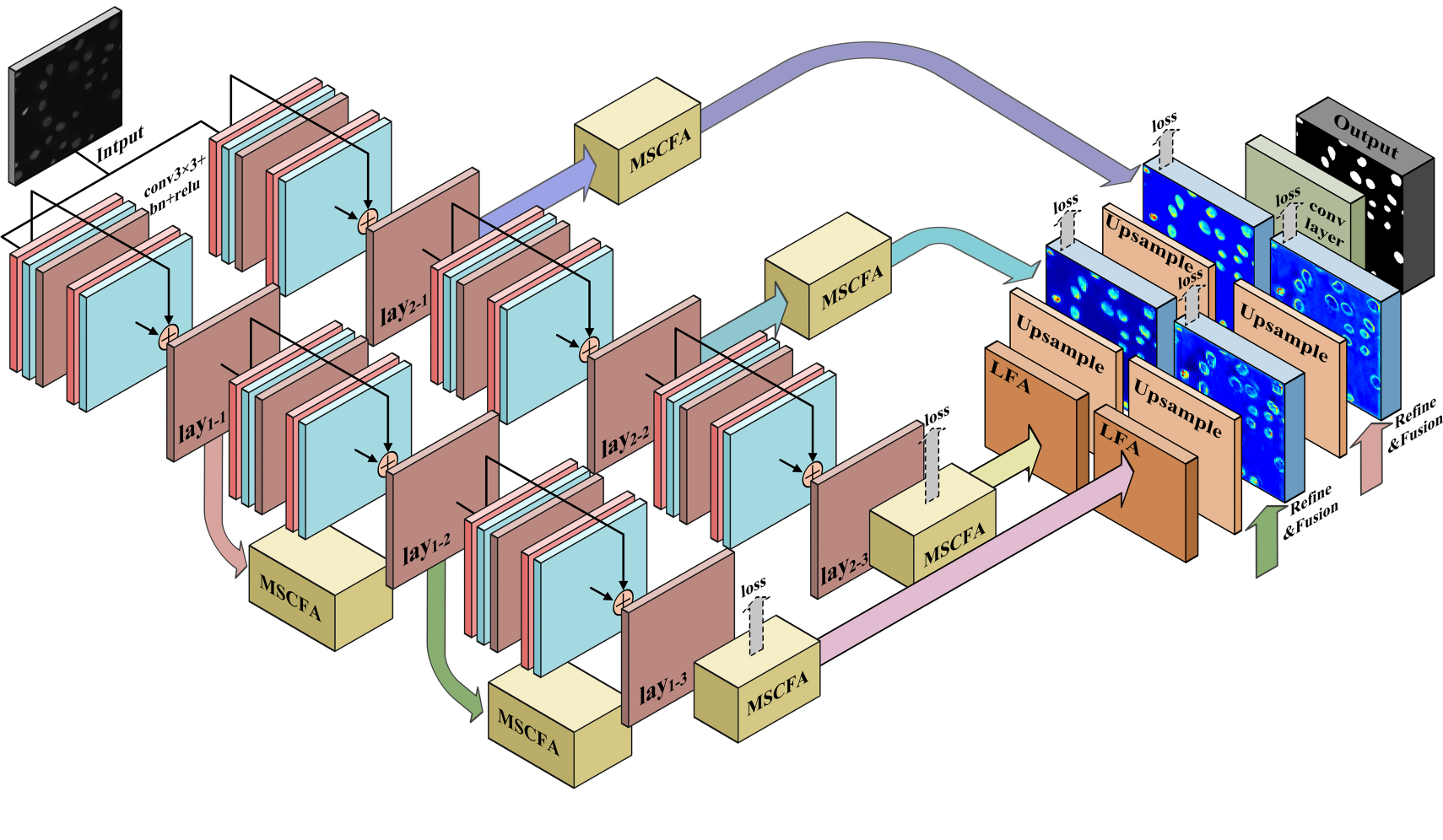}
	\caption{DmADs-Net main network diagram. Using $ResNet_{18}$ and $ResNet_{34}$ as the backbone network to complete feature extraction of different depths.In order to preserve more weak feature information, the MSCFA module is created and applied in skip connections and deep networks.LFA is created to enhance the contextual association of high-level semantic information. FRFB is created to complete the fusion of different information. A deep supervision mechanism is used to train the reinforcement network.}
	\label{fig2}
\end{figure*}

By integrating shallow and deep features, FCN-8s can capture both the details and global context of images, thus maintaining good structural integrity. However, FCN-8s has shown certain limitations in medical image segmentation, as the model restores image resolution through multiple upsampling steps, which may lead to the loss of some detail information. Particularly in larger scale upsampling, the restoration of details might not be sufficiently precise, resulting in blurred edge regions. MultiResUNet\cite{51} has made significant structural improvements to U-Net by incorporating multi-scale convolution and deep feature fusion mechanisms, enhancing the model's capability to process medical images of various modalities. Although MultiResUNet has made progress in feature integration, it still faces limitations in further refining and utilizing deep semantic information. Such limitations could result in less flexibility in the semantic fusion of deep features, potentially affecting the accuracy and robustness of the final segmentation results.

CFNet\cite{52} integrates multi-view attention mechanisms and adaptive fusion strategies, leading to significant performance improvements in the domain of medical image segmentation. By utilizing cross-scale feature fusion and multi-view attention mechanisms, CFNet effectively captures and integrates features across different scales, enhancing the recognition of image details and contextual information. However, CFNet's reliance on complex attention mechanisms may lead to overfitting of certain features, affecting its generalization performance. Despite the use of an adaptive fusion weight strategy to bridge semantic gaps across layers, the semantic disparity in highly heterogeneous medical images may still result in insufficient feature fusion.

Similarly, the KiU-Net model optimizes the capture of small structures and the precision of boundary segmentation by combining Kite-Net and U-Net. This design allows KiU-Net to extract detailed and boundary information in deep layers while maintaining the capture of high-level and low-level features, thereby improving segmentation accuracy. Nevertheless, KiU-Net has limitations in its feature fusion strategy, particularly when dealing with images of complex backgrounds and structures, where the feature fusion between Kite-Net and U-Net branches may not be smooth. Moreover, although the Kite-Net branch is sensitive to details, it may be limited in capturing deep, high-level features, affecting the understanding of the overall image context and complex structures.

TMAHU-Net\cite{53} innovatively introduces a multi-attention hybrid network that leverages the strengths of CNNs and Transformers, significantly enhancing the accuracy of skin lesion segmentation. This model utilizes a depth-wise separable convolutional attention module to flexibly allocate attention weights, thereby reinforcing the capture of channel and spatial information. While TMAHU-Net successfully integrates the advantages of CNNs and Transformers, the Transformer component, despite offering a global perspective, might be less intuitive in capturing local detail features, particularly in areas with fuzzy or highly irregular lesion boundaries.

On the other hand, the MSNet\cite{42}  adopts a multi-scale subtraction network structure to address the issue of redundancy in multi-layer feature fusion inherent in traditional U-shaped architectures, showing substantial benefits for precise polyp segmentation. However, while MSNet achieves impressive results in polyp segmentation, its multi-scale subtraction approach may lead to imbalanced feature capture, especially in scenarios with highly variable polyp sizes or complex structures.

Through detailed experimental analysis, we observed that current advanced medical image segmentation algorithms still have room for improvement in capturing lesion edge details and maintaining target integrity (as in Fig.~\ref{fig1}). This phenomenon indicates that existing methods lack efficiency in capturing and utilizing lesion texture information during the feature extraction phase and possess limited capability in utilizing and integrating multi-semantic information that includes the target during the feature reconstruction stage.

Addressing these challenges, this study introduces a Dense Multiscale Attention Network operating under a Deep Supervision mechanism (as in Fig.~\ref{fig2}). To enhance the expression of texture information, we developed the Multi-scale Convolutional Feature Attention Block (MSCFA). MSCFA employs an innovative multi-scale feature processing strategy that aggregates features using multiple dense residual blocks and multi-path parallel convolutions, thereby expanding the receptive field and achieving broad feature correlation. To tackle the issue of information fusion in feature reconstruction, we further proposed the Feature Refinement and Fusion Block (FRFB), which enhances shallow network information through residual blocks and edge spatial attention mechanisms, converting deep semantic information into channel weights to guide the reconstruction of shallow features. Additionally, the Local Feature Attention Block (LFA) utilizes a block processing strategy to enhance deep semantic information, establishing more extensive feature associations. A deep supervision mechanism is applied in the feature reconstruction phase, providing extra supervision to the reconstruction process and further optimizing model performance.

In this paper, we contribute as follows:

(1) Introduce the Multi-scale Convolutional Feature Attention Block, which enhances the expression of detail information and establishes feature correlations across a broader area through dense residual blocks combined with an expanded convolutional receptive field.

(2) Based on the block processing concept, the Local Feature Attention Block is proposed. This module reallocates channel weights for high-level semantic information, emphasizing target features, and then enhances local semantic associations through block processing.

(3) During the feature reconstruction phase, the Feature Refinement and Fusion Block is introduced to merge semantic information from different layers. This block recovers detail information through edge spatial attention mechanisms and performs feature compensation with residual connections, while also channel filtering deep information to highlight target features.

(4) To demonstrate the processing power of DmADs-Net, we chose five datasets of different sizes that are currently publicly available (ISIC2016\cite{14}, JSRT\cite{15}, Glas\cite{16}, DSB2018\cite{12}, BUSI\cite{17}) to continue our comparison experiments. The combined prediction results and metrics from the results show that DmADs-Net shows superior results in the localisation and characterisation of lesion targets.

\section{Related Work}

In this section we give a brief overview of the development of deep learning in the field of medical image segmentation, a brief introduction to deep supervisory mechanisms and finally we outline attention-related content.

\subsection{Medical image segmentation and deep learning}

Since the beginning of FCN, excellent networks in the field of medical image segmentation have been constantly updated, with SegNet introducing the "encoder-decoder" structure for the first time, and U-Net bringing the "encoder-decoder" structure to the forefront of medical image segmentation. Unlike previous networks, U-Net uses a form of stitching when fusing features of skip connection in order to be able to retain more dimensional information. Although the dimensional stitching format has some advantages in semantic segmentation tasks, the thicker features also increase memory consumption. Each layer of the U-Net encoder is passed only to the corresponding decoder, and this processing may be subject to feature discrepancy phenomena. UNet++ enhances the processing of skip connection on top of U-Net. UNet++ passes deeper features through the skip connection to the shallow processing of the decoder, thus reducing the possible semantic gap problem. In addition, a deep supervision mechanism is used in UNet++ to enhance the supervision of the training process. 

Dual Encoder\cite{18} uses a pre-processing approach to enhance the focus on feature information, and stained cell section samples are pre-processed to enhance the expression of edge features in the sample. The pre-processed feature information is then fed into the network together with the original image for further feature extraction, and in each layer, the fusion of the two-way feature information is reinforced by the constructed Attention Skip Module and passed to the decoder for feature stitching. The Compact Split-Attention block (CSA) is created in DCSAU-Net\cite{19} for multi-scale feature extraction, and channel attention\cite{20} is also added to the CSA to enhance the focus on weights. To enhance the focus on feature information, CSA is applied not only to each layer of the encoder stage, but also to the decoder stage, where feature information is recaptured after dimensional stitching. In order to process 3D medical image data, a new densely connected U-Net is proposed in H-DenseUNet\cite{21}. First the network converts the 3D image information into 2D features and sends them to 2DDenseUNet to extract the features of each layer slice, then the stitched together features are sent to 3DDenseUNet together with the original 3D image to complete the feature fusion. The brain tissue is relatively dense, and the lesions that occur here are more challenging to segment compared to other locations. In order to enhance the detection ability of brain tumors and improve the accuracy of medical diagnosis, Li S\cite{22} and others proposed a new region-of-interest-aided (ROI-aided) segmentation technique for the diagnosis of brain tumors. To avoid the influence of information from other human tissues on the extraction of tumor features, the network first uses a 2D U-Net for localization and then uses a 3D U-Net for binary segmentation. This network has achieved effective results.

The Self-attention proposed in Transformer\cite{23} has shown excellent performance in semantic segmentation, and in recent years Transformer has also been gradually used in the field of medical image segmentation. Since CNNs are limited by the convolutional kernel when performing feature extraction, they are unable to obtain global association information. TransUNet\cite{24} combines Transformer with U-Net, first using CNN to complete shallow feature extraction, and subsequently fusing features in Transformer for global association, thus improving the network's focus on global information. Considering that the number of samples in medical image datasets is generally small, and that it is often not optimal to train a network containing a Transformer, MedT\cite{25} proposed a "local-global" training strategy. MedT enhances the learning of local feature information by chunking the samples, while increasing the amount of data and improving the overall performance of the network. HmsU-Net\cite{54} showcased its innovative capability in addressing multi-scale medical image segmentation challenges by integrating CNN and Transformer technologies. The model adopted a parallel architecture, effectively merging local details with global contextual features, and resolved the consistency issues between the two feature extraction mechanisms through a multi-scale feature fusion module. The high performance of HmsU-Net in precise medical image segmentation allowed it to surpass existing advanced technologies across multiple evaluation metrics.

As medical technology continues to improve, the medical images that can be acquired become more abundant and varied; the need for high quality diagnosis relies on more accurate medical image segmentation techniques. Although many excellent algorithms have emerged in the field of medical image segmentation, there are still significant challenges in the field of medical image segmentation in the face of a more diverse and complex sample environment. We therefore propose DmADs-Net, which is expected to enable the task of segmentation of different kinds of medical image datasets and to improve the focus on feature information.

\subsection{Deep supervision}

A deep supervision mechanism is proposed in DSN\cite{26}, which is mainly used to solve the problem of gradient disappearance and gradient explosion that occurs during the training process of the network, to improve the transparency of the intermediate layers during the training process, and to enhance the learning ability of the network training process for features. The main idea of deep supervision is to add branch outputs directly to the intermediate layer and supervise the branch output results, setting the accompanying loss function. The main idea of deep supervision is to add branch outputs directly to the intermediate layer and supervise the branch output results, setting the accompanying loss function.

Due to the good performance of deep supervision mechanisms, their ideas are used in various areas of computer vision. In the field of Building Change Detection (BCD), DSA-Net\cite{27} adds branching outputs to each layer of the decoder stage in order to enhance the feature representation capability of the intermediate layers of the network and strengthen the representation of building features, and calculates the loss that accompanies the output results, thus further guiding the network training. In the medical field, for accurate localization of breast lumps as well as lesion locations, Rajalakshmi N R\cite{28} et al. have achieved excellent results by optimizing the learning process using deep supervised mechanisms on top of U-Net.

The same deep supervision mechanism is used in our network (as in Fig.~\ref{fig2}), with branch outputs in each layer of the feature fusion phase. Considering the importance of high-level semantic information, we also set branching outputs for the MSCFA of the bottleneck layer to enhance the learning supervision for the bottleneck layer. In addition, considering that too many loss functions may cause the network to converge slowly, we add weights to all the accompanying loss functions to control the proportion of accompanying losses in the total losses. The details of the loss function are described in detail in Part IV of the experiments.

\subsection{Attention mechanism}

As deep learning develops more deeply, the feature extraction capabilities of networks are receiving wider attention. Since the feature extraction capability of CNN is limited by the convolutional kernel, networks using traditional convolution can only focus on local information and it is difficult to capture feature information from a global perspective. Therefore, in order to gain attention to global features and improve the dependencies between remote features, SENet\cite{20}, Non-local Attention\cite{29}, SKNet\cite{30} and other networks have appeared one after another. By weighting each pixel in the form of calculating weights from the global level, the strength of the feature information is improved and the overall performance of the network is enhanced.

In recent years, attention mechanisms have played an equally important role in various areas of computer vision. Consider the fact that the structure of the attention module cannot be modified manually during the training process. $A^2 N$\cite{31} proposed an Attention in Attention Block ($A^2 B$) using Dynamic Convolution\cite{32} in order to implement single image super-resolution (SISR). The module adapts the weights to the input and performs a discard operation on unimportant features, thus streamlining the number of parameters and improving the focus on important features. To achieve the fusion of infrared and visible images, Li Y et al.\cite{33} proposed a feature extraction module based on the Laplace gradient operator and enhanced attention to depth features by ECA\cite{34}. The features of the two different images are finally fused together through multi-scale feature extraction and feature noticing.

In the field of segmentation, in order to obtain finer segmented images of urban buildings, MANet\cite{35} used softmax to pre-process the features, reducing the computational complexity, and named the module Kernel attention mechanism (KAM). In MANet, KAM and ECA process the same feature information in parallel to enhance the features in terms of both position dependence and channel dependence, respectively, with good results. In order to improve the accuracy of vessel localization, RADCU-Net\cite{36} utilized residual attention to construct a dual-supervision cascaded network, which demonstrated good performance in retinal segmentation tasks.

The segmentation targets faced in the field of medical image segmentation often have a high degree of similarity or identical texture to their surroundings. To obtain finer edge and detail information, increased attention to weak features around the target is required. We have therefore created MSCFA and LFA to complete the feature extraction of medical images, the details of which are described in the third part.

\section{Method}

In this section, we introduce the main architecture of DmADs-Net: Multi-scale Convolutional Feature Attention Block (MSCFA), Local Feature Attention Block (LFAB), Feature Refinement and Fusion Block (FRFB), in turn, and finally show the algorithmic flow of DmADs-Net, as in Algorithm 1.

\subsection{Network structure}

Considering that the feature information captured by the network structure at different depths is not the same. Therefore, in DmADs-Net, $ResNet_{18}$, $ResNet_{34}$ are used as the backbone network in order to extract feature information at different depths more efficiently and to prevent the problem of gradient disappearance due to increasing depth. It is worth noting that since a large number of features in medical images have the same texture features as their surroundings and are relatively weak, we only kept the convolutional layer in the middle part of ResNet\cite{37} and discarded operations such as pooling to retain richer feature information for the network to process.

In order to obtain feature information at different depths and with different sensory fields, each layer of processing in the backbone network changes the size of the feature map. This operation will inevitably result in loss of features. In order to complete the supplement of feature information, we set up skip connections for the middle two layers of processing to pass low-level semantic information to the feature fusion stage.

For the processing of skip connections, MSCFA is created to reinforce the feature strength of low-level semantic information. In addition, MSCFA is applied to the processing of depth features in addition to skip connections, thus preserving a richer set of features for processing in the bottleneck layer. Deeper features contain more semantic information and in the bottleneck layer, LFA is created in order to reinforce the dependencies between high-level semantic information. In LFA, the feature information is first enhanced from a global perspective by the SE block proposed by SENet, and then the high-level semantic information is dependent on the association in the form of multi-scale chunking.

The feature fusion phase needs to accept not only low-level semantic information from the skip connection, but also high-level semantic information from the bottleneck layer. How to fuse these two features together more effectively becomes the focus of the research, hence the creation of FRFB in the fusion phase. FRFB is primarily a residual structure that enhances the granularity of different feature information through branching, ultimately achieving better fusion of feature information and outputting a final segmentation prediction map.

In order to increase the transparency of the intermediate layers during training and to further optimise the operation of the network, DmADs-Net employs a deep supervision mechanism. As shown in Fig.~\ref{fig2}, we set up depth supervision at the fusion stage as well as at the bottleneck layer to compute the loss of branch output features with the same size of GT.

\subsection{Multi-scale Convolutional Feature Attention Block}

\begin{figure}
	\centering
	\includegraphics[scale=1.4]{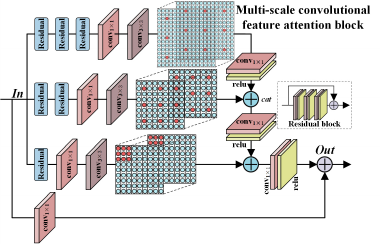}
	\caption{Multi-scale Convolutional Feature Attention Block.}
	\label{fig3}
\end{figure}

Residual blocks, through skip connections, focus on detail information while enabling feature reuse. Moreover, chaining multiple residual blocks enhances the network's ability to learn more complex and abstract features, as demonstrated by past work\cite{37,44,45,46}. The advantages of residual blocks have been well-established. ASPP\cite{47} captures multi-scale feature information through convolutions with varying dilation rates. By expanding the receptive field, the model effectively establishes broader associations among features. Inspired by these insights, we propose the Multi-scale Convolutional Feature Attention Block (MSCFA), as in Fig.~\ref{fig3}.

In the network, it enhances the features of skip connections and bottleneck layers. The whole module is designed on a residual structure and three paths with different sensory fields are constructed. A Residual block consisting of a dense 3$\times$3 convolution is first constructed and used in an iterative manner to focus the feature information. Subsequently, the feature information is associated by means of 3$\times$3 convolutions with different sensory fields.

In the upper path processing (as in Eq.~\ref{eq:1}), since the largest perceptual field expansion is used, we first use a dense Residual block to increase the strength of important features, and finally use the convolution of the expanded perceptual field to enhance the remote dependencies. Similarly, the same treatment is used in the middle (as in Eq.~\ref{eq:2}) and lower path treatment (as in Eq.~\ref{eq:3}), with the difference that the number of Residual blocks is adjusted accordingly as the sensory field changes.

Since the feature information captured by the three processes is different, we fuse the feature information in the form of channel stitching and integrate the feature information channel by channel using 1×1 convolution in order to maximise the integrity of the feature information (as in Eq.~\ref{eq:4}). Finally, we complement the original features by summing the features (as in Eq.~\ref{eq:5}). Where $f_{3\times 3}\left( \cdot \right)$ is the 3$\times$3 convolution, $f_{1\times 1}\left( \cdot \right)$ is the 1$\times$1 convolution, $F_{res}\left( \cdot \right) $ is the Residual block and $\alpha $ is the $\text{Re}LU$.

\begin{figure}
	\centering
	\includegraphics[scale=0.29]{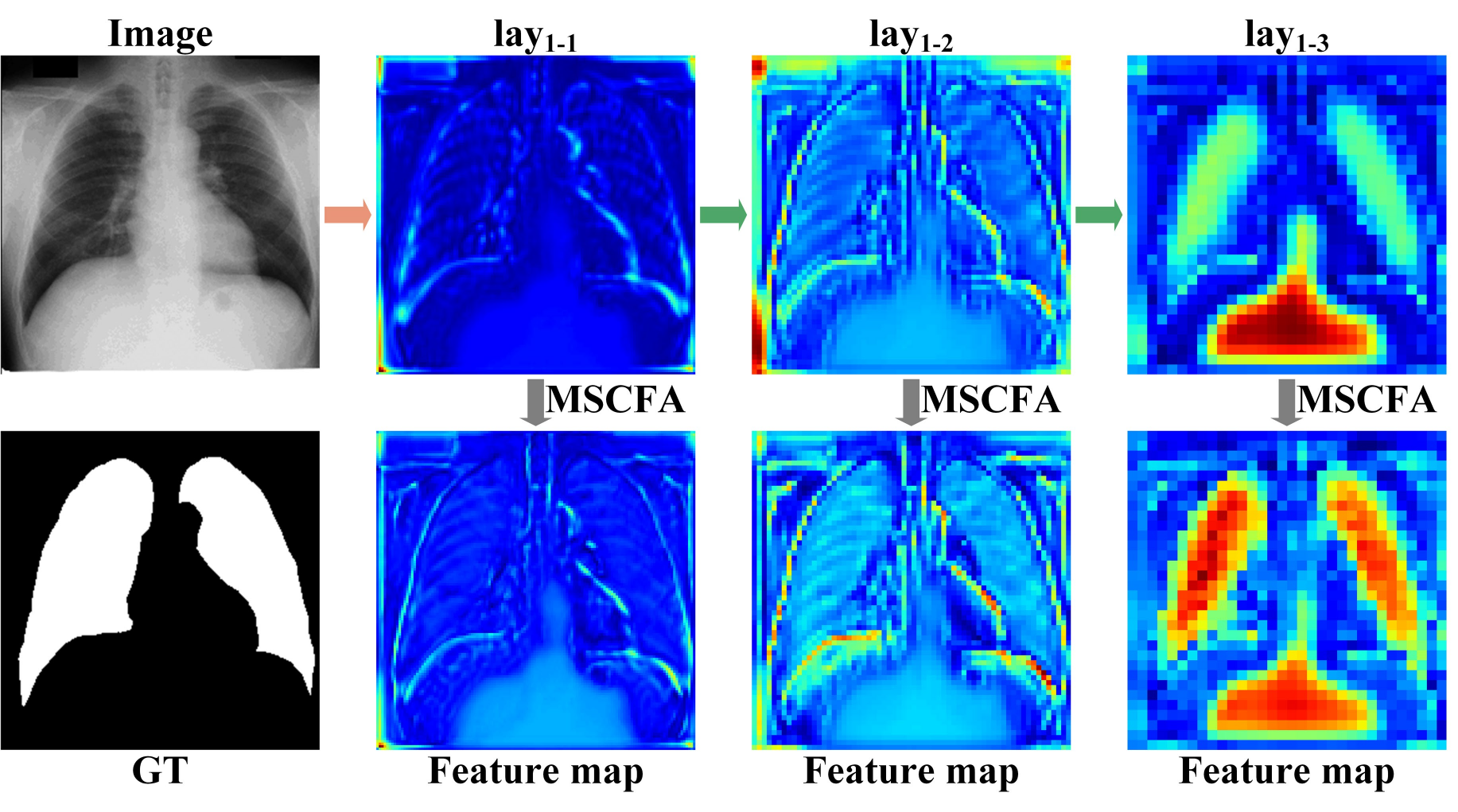}
	\caption{Feature attention map after MSCFA processing.}
	\label{fig4}
\end{figure}

As in Fig.~\ref{fig4}, the processing results of MSCFA at different stages are displayed. Within the three layers of semantic information captured by the backbone network ($lay_{1-i}$), the first two layers contain a large amount of target texture information, while the last layer is rich in macro-location information of both the target and its surroundings. After processing through dense residual blocks and parallel convolutions with varying scale receptive fields, the texture and edge information in the first two layers' features are significantly enhanced. In the third layer, the feature target location weight captured by the backbone network is lower, while the environmental weight is higher. After processing with convolutions that expand the receptive field, a broader association among feature information is established, allowing the network to more effectively recognize target features, ultimately enhancing the weights of both lung sides. Consequently, the corresponding environmental weight is significantly reduced.

\begin{equation}
	\label{eq:1}
	Out_1=f_{3\times 3}^{rate=4}\left( F_{res}\left( F_{res}\left( F_{res}\left( In \right) \right) \right) \right)
\end{equation}
\begin{equation}
	\label{eq:2}
	Out_2=f_{3\times 3}^{rate=2}\left( F_{res}\left( F_{res}\left( In \right) \right) \right) 
\end{equation}
\begin{equation}
	\label{eq:3}
	Out_3=f_{3\times 3}^{rate=1}\left( F_{res}\left( In \right) \right) 
\end{equation}
\begin{equation}
	\label{eq:4}
	Out_{1-3}=\alpha f_{1\times 1}\left( \alpha f_{1\times 1}\left( Out_1,Out_2 \right) ,Out_3 \right) 
\end{equation}
\begin{equation}
	\label{eq:5}
	Out=In+\alpha f_{1\times 1}\left( Out_{1-3} \right) 
\end{equation}

\subsection{Local Feature Attention Block}

\begin{figure}
	\centering
	\includegraphics[scale=1.5]{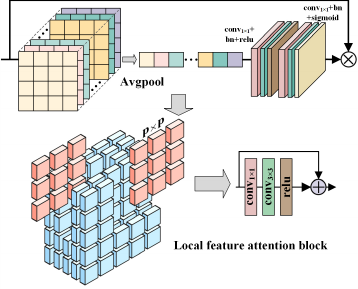}
	\caption{Local Feature Attention Block.}
	\label{fig5}
\end{figure}

Due to the high sensitivity of channel attention to the channels of potential target features, we aim to utilize it to enhance the intensity of the target's channels in deep semantic information. Additionally, influenced by the concept of feature patch processing \cite{48,49,50}, we construct a feature slicing processing scheme to improve the model's learning of feature information.

Fig.~\ref{fig5} displays the Local Feature Attention Block (LFA), which serves to strengthen the network's focus on high-level semantic information. High-level semantic information contains a large number of semantic features, and traditional convolution is only able to establish associations with local features within the convolution kernel during processing, without being able to focus on contextual features from a global perspective. Therefore, in the LFA, we first compress the feature information of $X\in \mathbb{R}^{H\times W\times C}$ using the SE block to obtain the global feature information $z\in \mathbb{R}^{1\times 1\times C}$ (as in Eq.~\ref{eq:6}). To enable the network to learn to use global information and to learn to selectively emphasise feature information, highlighting useful features and suppressing unimportant features, a gating mechanism is used (as in Eq.~\ref{eq:7}), where $W_1=\mathbb{R}^{\frac{C}{2}\times C}$, $W_2=\mathbb{R}^{C\times \frac{C}{2}}$, $\delta$ is the $sigmod$ and $\alpha$ is $\text{Re}LU$, to obtain the feature $S$. Subsequent channel multiplication with the original features embeds the global information into the original feature information (as in Eq.~\ref{eq:8}) to obtain feature $O$. Subsequently, we sliced the feature $O$ into multiple chunks $o_i\in \mathbb{R}^{p\times p}$ according to $p\times p$ size and performed 3$\times$3 convolution on each of the sliced chunks to strengthen the association of regional feature information (as in Eq.~\ref{eq:9}), where $n=\lceil W/p \rceil$. We set up ratios $p$ of different sizes to correlate regions of feature information in a parallel way (as in Eq.\ref{eq:10}), where $O_i$ is the result of the different ratios $p$.

\begin{figure}
	\centering
	\includegraphics[scale=0.35]{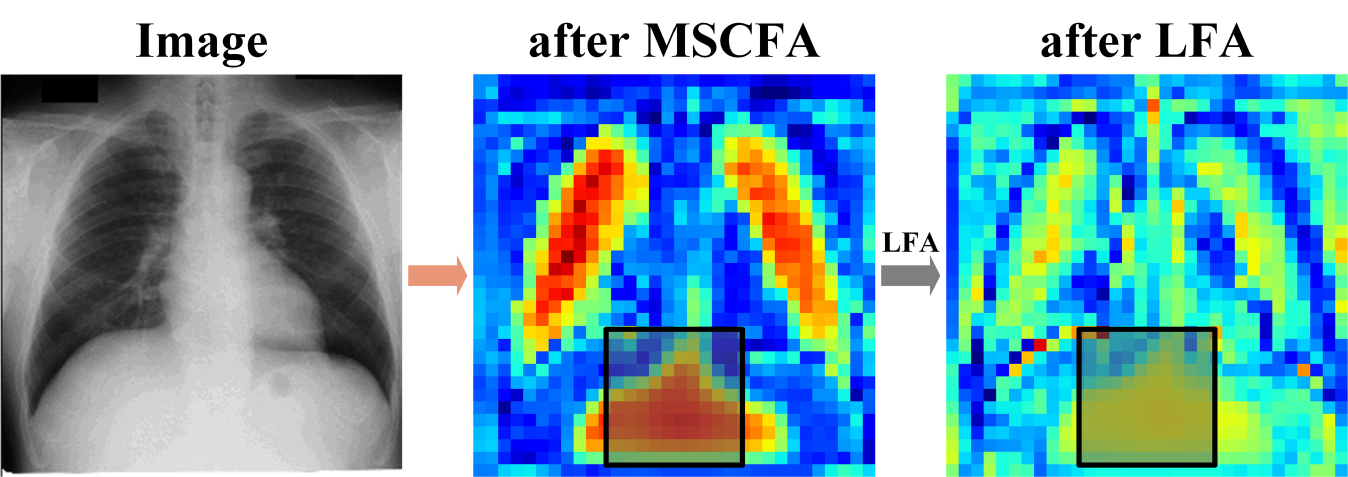}
	\caption{Feature attention map after LFA processing.}
	\label{fig6}
\end{figure}

Fig.~\ref{fig6} displays the feature maps after processing with the LFA. It is clearly observable that all weights are reduced. This reduction is mainly due to the lung and its surrounding environment sharing similar texture features, leading the channel attention to adjust the weights of channels with similar information during processing. However, the texture information of the lung target is still enhanced. This enhancement is mainly due to the patch processing approach, which allows deep information to establish tighter local semantic connections. As a result, the position weight indicated by the gray box in the image is effectively reduced. Following LFA processing, the continuity of edge features is significantly improved, playing a crucial role in guiding the subsequent lesion target reconstruction.

\begin{equation}
	\label{eq:6}
	z=\frac{1}{H\times W}\sum_{i=1}^H{\sum_{j=1}^W{X\left( i,j \right)}}
\end{equation}
\begin{equation}
	\label{eq:7}
	S=\delta \left( W_2\alpha \left( W_1z \right) \right)  
\end{equation}
\begin{equation}
	\label{eq:8}
	O=X\otimes S 
\end{equation}
\begin{equation}
	\label{eq:9}
	O=\alpha f_{1\times 1}\left( f_{3\times 3}\left( \left[ o_1,...o_n \right] \right) \right)  
\end{equation}
\begin{equation}
	\label{eq:10}
	Out=\alpha f_{1\times 1}\left( O_1,O_2,O_3,O_4 \right)  
\end{equation}

\subsection{Feature Refinement and Fusion Block}

Shallow networks extract features that contain more texture and detail information, while deep networks capture more abstract information, including scene context and macro-location information of objects. To enhance detail information, it is necessary to focus on weak features, and residual blocks have been proven effective in past studies. Through extensive experimentation by various researchers, ESA\cite{38} has been shown to significantly focus on edge and texture information. Therefore, we aim to utilize a combination of both to enhance the edge intensity of lesion target information in the features captured by shallow networks, allowing the network to achieve a more complete segmentation result. To maintain high sensitivity to channel information, the features obtained by deep networks are updated through channel attention. The fusion and interaction of these two different features ultimately strengthen the lesion target information.

\begin{figure}
	\centering
	\includegraphics[scale=1.3]{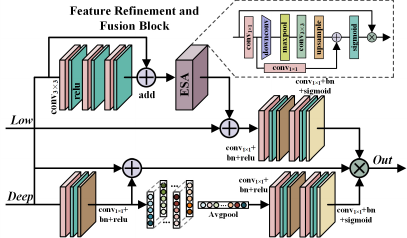}
	\caption{Feature Refinement and Fusion Block.}
	\label{fig7}
\end{figure}

Therefore, in the feature fusion stage, to more effectively enhance edge feature information and strengthen the fusion of features of different scales, we propose the Feature Refinement and Fusion Block (FRFB), as in Fig.~\ref{fig7}. The FRFB receives two parameters and the residual structure is still used in the first path processing when processing low-level semantic information. The feature information is first enhanced by dense blocks of residuals, and subsequently we introduce ESA to capture the association between more feature information and superimpose the refined feature information onto the original features. Finally, the feature information is adaptively adjusted by the gating mechanism (as in Eq.\ref{eq:11}) to obtain $Low_1$, where $F_{ESA}\left( \cdot \right) $ is the ESA and $F_{res}\left( \cdot \right) $ is the Residual block.

The second path low-level semantic information is fused with the reduced-dimensional high-level semantic information for features (as in Eq.~\ref{eq:12}) to obtain $Fu$. And after the high level semantic information is downscaled to the dimension where the low level semantic information is located, through the SE block, $Deep_2$ is obtained to participate directly in the final channel multiplication to complete the fusion process (as in Eq.~\ref{eq:13}) to obtain $Out$.

\begin{figure}
	\centering
	\includegraphics[scale=0.35]{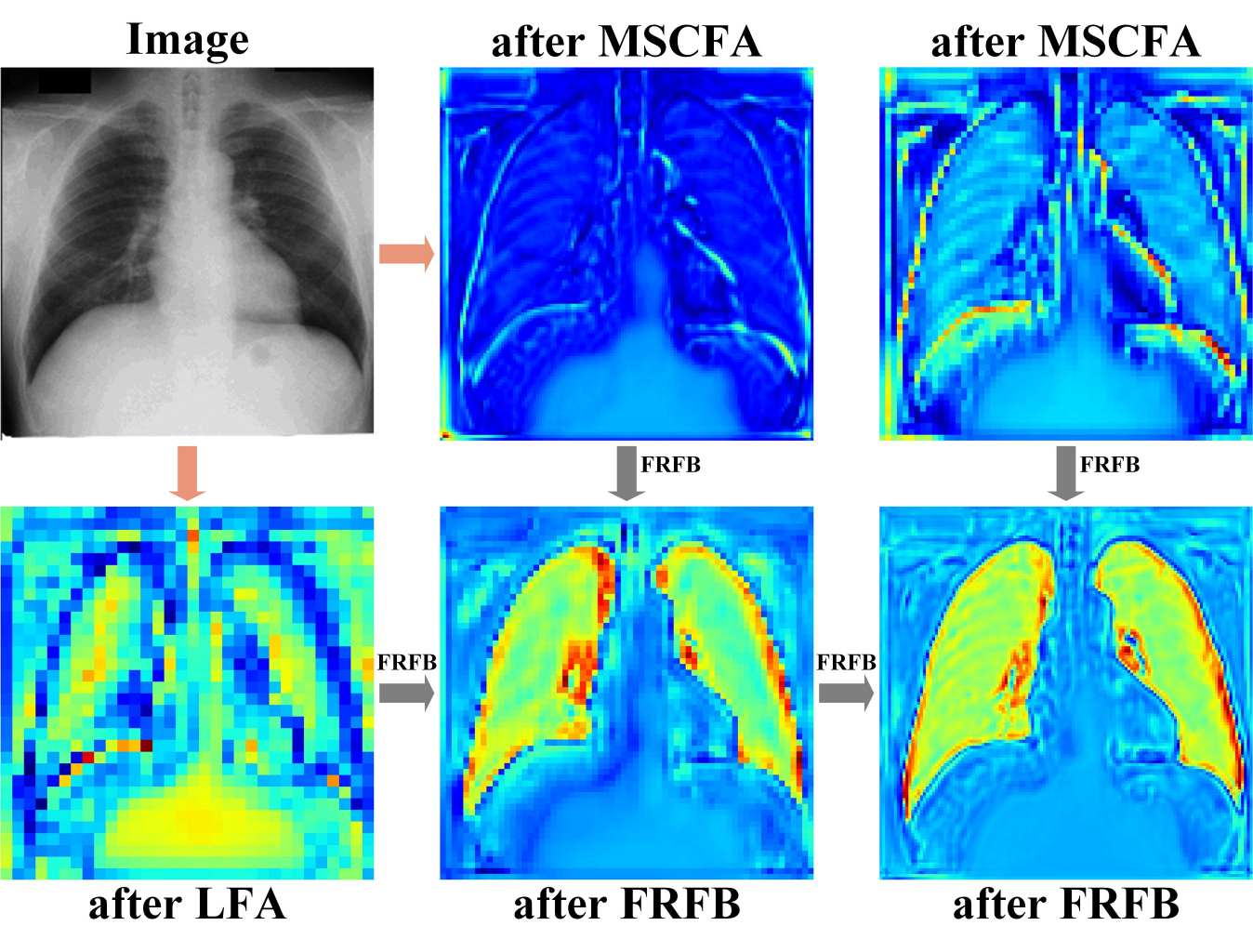}
	\caption{Feature attention map of high-level semantic information processed by LFA and low-level semantic information processed by MSCFA processed by FRFB.}
	\label{fig8}
\end{figure}

Fig.~\ref{fig8} displays the feature processing results on one path during the feature reconstruction phase. FRFB receives shallow features rich in detail and texture information processed by MSCFA, as well as deep features containing rich target macro-location information processed by LFA. After fusion and reconstruction, the edge information and integrity of the target are further enhanced. Moreover, compared to environmental information, target information maintains a higher weight. This is primarily due to the further enhancement of detail information from shallow networks and the refinement of channels containing target information in deep semantic information within FRFB.

\begin{equation}
	\label{eq:11}
	Low_1=\delta \left( W_2\alpha \left( W_1F_{ESA}\left( F_{res}\left( Low \right) \right) \right) \right) 
\end{equation}
\begin{equation}
	\label{eq:12}
	Fu=Low+Deep
\end{equation}
\begin{equation}
	\label{eq:13}
	Out=Low_1\otimes Fu\otimes Deep_2
\end{equation}

\subsection{Feature Visualization}

\begin{figure*}[!t]
	\centering
	\includegraphics[width=\linewidth]{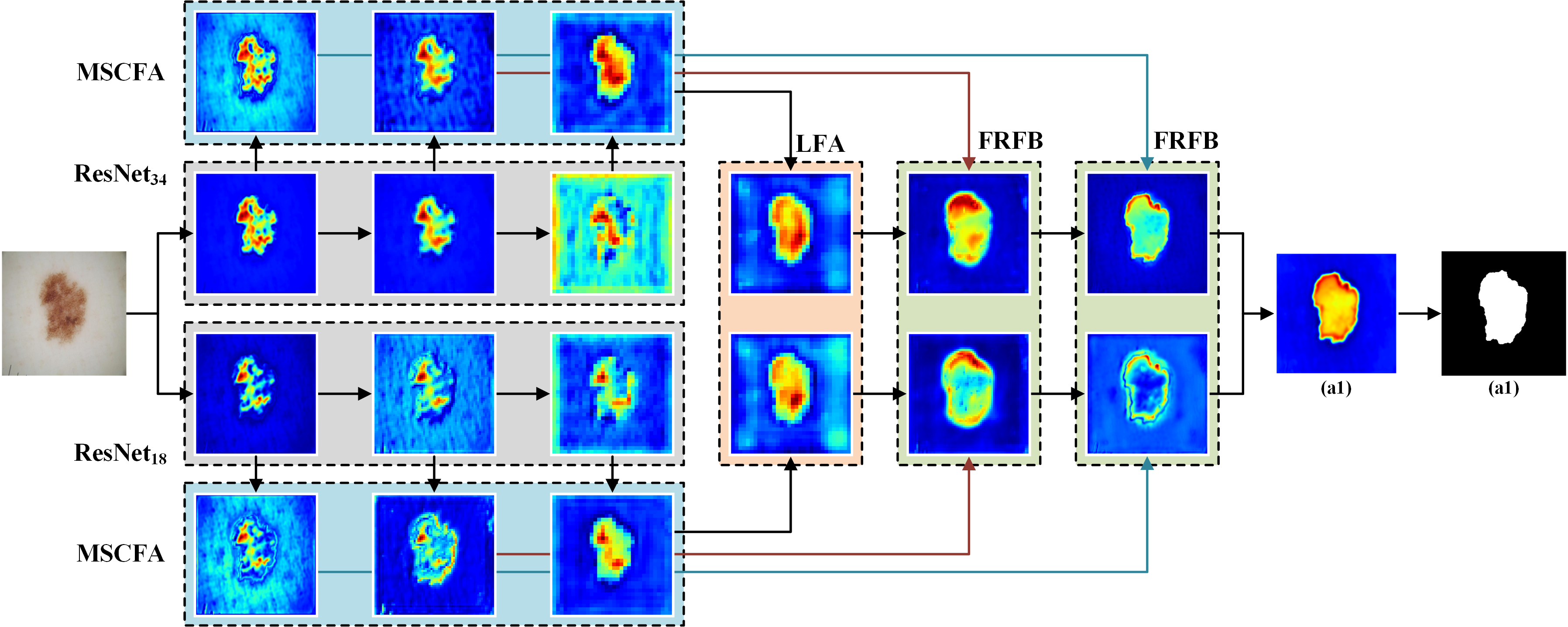}
	\caption{Demonstration of feature information acquired by DmADs-Net in different stages.}
	\label{fig29}
\end{figure*}

Fig.~\ref{fig29} shows a set of feature maps generated by the model when processing skin data samples, with different background colors indicating the main architecture of the model and arrows showing the flow direction of feature information. The gray background represents the information extracted by $ResNet_{34}$ and $ResNet_{18}$ at three different depths and scales, showing noticeable differences in the information they focus on, yet both annotate the features of the lesion area. After processing with MSCFA against a blue background, the information weight captured by the backbone network significantly increases, especially in terms of target integrity, closer to the lesion area, demonstrating the positive impact of MSCFA’s dense residual blocks and parallel multi-scale receptive field strategy on performance.

After LFA processing, particularly in the $ResNet_{18}$ path, the macro-localization of the target is effectively enhanced, proving the facilitative role of patch processing in modeling deep semantic information. In the subsequent feature reconstruction phase, FRFB integrates the results of LFA and MSCFA, and through two levels of FRFB processing, significantly enhances the edge information of the target. This further confirms that the residual blocks and ESA within FRFB enhance the model's sensitivity to edge and texture information. Moreover, the processing of deep semantic information by FRFB increases the weight of the target channel, leading to better model performance by utilizing macro-location information to guide target reconstruction.

\section{Experiment}

In this section, we will provide an overview of a series of experiments conducted on DmADs-Net. We will begin by describing the experimental environment and introducing the datasets used in the experiments. Then, we will present some results of the comparative experiments and analyze the findings. Finally, to further demonstrate the rationality of the network architecture, we will conduct ablative experiments on DmADs-Net.

\begin{algorithm}[H]
	\caption{DmADs-Net}\label{alg:alg1}
	\begin{algorithmic}
		\STATE 
		\STATE \textbf{Input:} $Image=X_i;~GT=G_i~i\in[1,n]$
		\STATE \textbf{Output:} $Segmentation\ Map\ S,Loss_1,Loss_2$
		\WHILE{$not~converge$}
		\STATE {\textsc{TRAIN}}$(\mathbf{X_i},\mathbf{G_i})$
		\STATE $list_1=\text{Re}sNet_{18}\left( X_i \right)$
		\STATE $list_2=\text{Re}sNet_{34}\left( X_i \right)$
		\STATE $Mlist_1=MSCFA\left( list_1 \right)$ 
		\STATE $Loss_1+=loss\left( Mlist_{\left( 1,3 \right)},G_i \right)$
		\STATE $Mlist_2=MSCFA\left( list_2 \right)$
		\STATE $Loss_1+=loss\left( Mlist_{\left( 2,3 \right)},G_i \right)$
		\STATE $Out_1=LFA\left( Mlist_{\left( 1,3 \right)} \right)$
		\STATE $Out_2=LFA\left( Mlist_{\left( 2,3 \right)} \right)$
		\hspace{0.5cm} \FOR{$k=0;$ $k<2;$ $k++;$}
		\STATE $Out_1=FRFB\left( Out_1\uparrow ,list_{\left( 1,k \right)} \right)$
		\STATE $Loss_1+=loss\left( Out_1,G_i \right)$
		\STATE $Out_2=FRFB\left( Out_2\uparrow ,list_{\left( 1,k \right)} \right)$
		\STATE $Loss_1+=loss\left( Out_2,G_i \right)$
		\ENDFOR
		\STATE $S=Fusion\left( Out_1,Out_2 \right)$
		\STATE $Loss_2=loss\left( S,G_i \right)$
		\STATE $Loss=\theta Loss_1+Loss_2$
		\STATE \textbf{Loss Function:}
		\STATE We use SoftIoU as our loss function to guide our training. We use the Adam optimizer to optimize our network.
		\ENDWHILE
	\end{algorithmic}
	\label{alg1}
\end{algorithm}

\subsection{Experimental parameters}

The training was performed using NVIDIA TITAN RTX GPUs, running on a Linux environment with Pytorch version 1.4.0. The maximum iteration for all compared methods was set to 400 rounds, with the best model compared every 10 rounds. If the best model did not update for 50 consecutive rounds, the training process was automatically stopped. Additionally, the same training strategy was applied to all compared methods on all datasets.

In medical image segmentation tasks, the Dice coefficient (Eq.~\ref{eq:14}) significantly represents the overlap between the predicted segmentation area and the actual annotated area, which is crucial for formulating subsequent diagnostic and treatment strategies. A high Dice coefficient indicates the model's high accuracy in identifying specific tissues or lesions, essential for evaluating the model's performance in maintaining pathological characteristics. The IoU-score (Eq.~\ref{eq:15}) also serves as a benchmark metric for assessing image segmentation quality, measuring the ratio of the intersection to the union between predicted and actual segmentations. In the context of medical imaging, the IoU score enables medical professionals to accurately assess the model's ability to differentiate between lesion and normal tissues, crucial for ensuring the clinical applicability of the model.

The Precision metric (Eq.~\ref{eq:16}) measures the model's ability to correctly identify pixels as lesions (positive class). High precision in the field of medical image segmentation reduces the occurrence of false positives, thereby avoiding misdiagnosis and unnecessary treatment for patients, making precision a key indicator of model performance, especially in medical applications where diagnostic sensitivity is critical. Similarly, Recall (Eq.~\ref{eq:17}) reflects the model's ability to capture the actual lesion area, where a high recall rate ensures that critical pathological information is not overlooked, crucial for preventing missed diagnoses and ensuring timely and accurate treatment for patients. During the disease assessment phase, a high recall rate aids physicians in comprehensive disease evaluation, leading to more precise treatment planning. Therefore, these four metrics provide a comprehensive and accurate set of performance evaluation tools for medical image segmentation, which we use as objective evaluation criteria in our experiments.

\begin{equation}
	\label{eq:14}
	Dice=\frac{2TP}{2TP+FP+FN}
\end{equation}
\begin{equation}
	\label{eq:15}
	IoU=\frac{TP}{FP+FN+TP}
\end{equation}
\begin{equation}
	\label{eq:16}
	Precision=\frac{TP}{TP+FP}
\end{equation}
\begin{equation}
	\label{eq:17}
	Recall=\frac{TP}{TP+FN}
\end{equation}

Binary cross-entropy loss, as in Eq.~\ref{eq:18}, is the most commonly used loss function in the field of medical image segmentation; therefore, we also chose it as the loss function for our experiments. The network employs a deep supervision mechanism, setting companion outputs at different stages and calculating the loss (Eq.~\ref{eq:19}). Finally, by adjusting the weights, the proportion of companion loss in the total loss (Eq.~\ref{eq:20}) is modified, where $L_{CE}\left( \cdot \right)$ represents the loss of the final output feature map. Drawing on past experimental experience, we set the value of $\theta $ to 0.5.

\begin{equation}
	\label{eq:18}
	\begin{aligned}
		L_{CE\left( p,\hat{p} \right)}=&-\frac{1}{WH}\sum_{x=0}^{W-1}{\sum_{y=0}^{H-1}{p\left( x,y \right) \log \hat{p}\left( x,y \right) }} \\
		&{{ +\left( 1-p\left( x,y \right) \right) \log \left( 1-\hat{p}\left( x,y \right) \right)}}
	\end{aligned}
\end{equation}
\begin{equation}
	\label{eq:19}
	L_{deep}=\sum_{i=1}^{n=6}{L_{IoU}\left( X,GT \right)}
\end{equation}
\begin{equation}
	\label{eq:20}
	L_{total}=L_{IoU}+\theta L_{deep}
\end{equation}

\subsection{Experimental dataset}

\subsubsection{JSRT}

The JSRT dataset consists of human chest images obtained through X-ray scanning, containing a large number of lung images with each sample having a resolution of 256$\times$256 pixels. In these lung X-ray images, the lung contours are often obscured by neighboring organs such as the heart, stomach, and spleen, leading to incomplete lung contour information. Compared to optical imaging, X-ray imaging has limitations in depicting edge details, often failing to capture the target's edge features accurately, resulting in lung images with relatively high blurriness of edge information. Given these characteristics, we choose the JSRT dataset for our experiments to test the model's performance in capturing edge information, especially its ability to handle blurred edges caused by anatomical structure occlusion and imaging technique limitations.

\subsubsection{ISIC2016}

The ISIC2016 dataset is specifically designed for the detection of skin lesion tissues, containing a rich collection of skin lesion samples presented at a resolution of 256$\times$256 pixels. This dataset emphasizes the diversity and complexity of samples, providing a broad range of cases for skin lesion detection. One challenge within the samples is that environmental factors, such as human hair, can obscure lesion areas, posing a severe test to the model's recognition capabilities. Moreover, since skin surface cells and lesion cells share a similar physiological environment, some samples exhibit a high visual similarity between lesion and non-lesion areas, further demanding the model's excellent discrimination ability. Given these characteristics, we use the ISIC2016 dataset as an evaluation tool to examine the model's detailed recognition and performance in handling skin lesion detection.

\subsubsection{DSB2018}

The DSB2018 dataset encompasses a wide range of cell imaging samples, with each sample having a resolution of 256$\times$256 pixels. This dataset is rich in features, containing cell samples captured in various lighting conditions, demonstrating the diversity in cell size, density, and morphology, all of which significantly challenge the model's generalization ability and adaptability. In stained scenarios, the uneven distribution of staining agents and the death of some cells can introduce significant interference noise in the samples, increasing the complexity of image analysis. In unstained dark-field scenarios, the visual information of cell edges is often unclear, thereby increasing the difficulty of model recognition. Given these characteristics and challenges, we choose the DSB2018 dataset for a comprehensive evaluation of model performance, ensuring the developed model can effectively handle complex biological image data in practical applications.

\subsubsection{BUSI}

The BUSI dataset is created for the treatment of breast cancer, a disease with a high mortality rate in women. The samples in the dataset are taken from 600 female patients at the location of the breast and classified into three categories: normal, benign and malignant. We chose to use this dataset to detect only the malignant category, and adjusted the sample size to 256$\times$256. The targets in this dataset have a high degree of similarity to their surroundings and therefore place high demands on the performance of the network.

\subsubsection{GlaS}

The GlaS dataset contains a large number of samples of cell sections from colon lesions, and this dataset is also characterised by sample diversity. To test the ability of DmADs-Net to handle weak feature information and dense features, the sample size of the GlaS dataset was adjusted to 128$\times$128.

\subsection{Comparative experiment}

Next, we outline our comparative experiments on five datasets, for each of which we carry out a subjective analysis with the help of result plots and select some of the results for local feature comparisons, followed by an objective analysis by means of various indicators.

\subsubsection{JSRT}

\begin{figure*}
	\centering
	\includegraphics[scale=0.20]{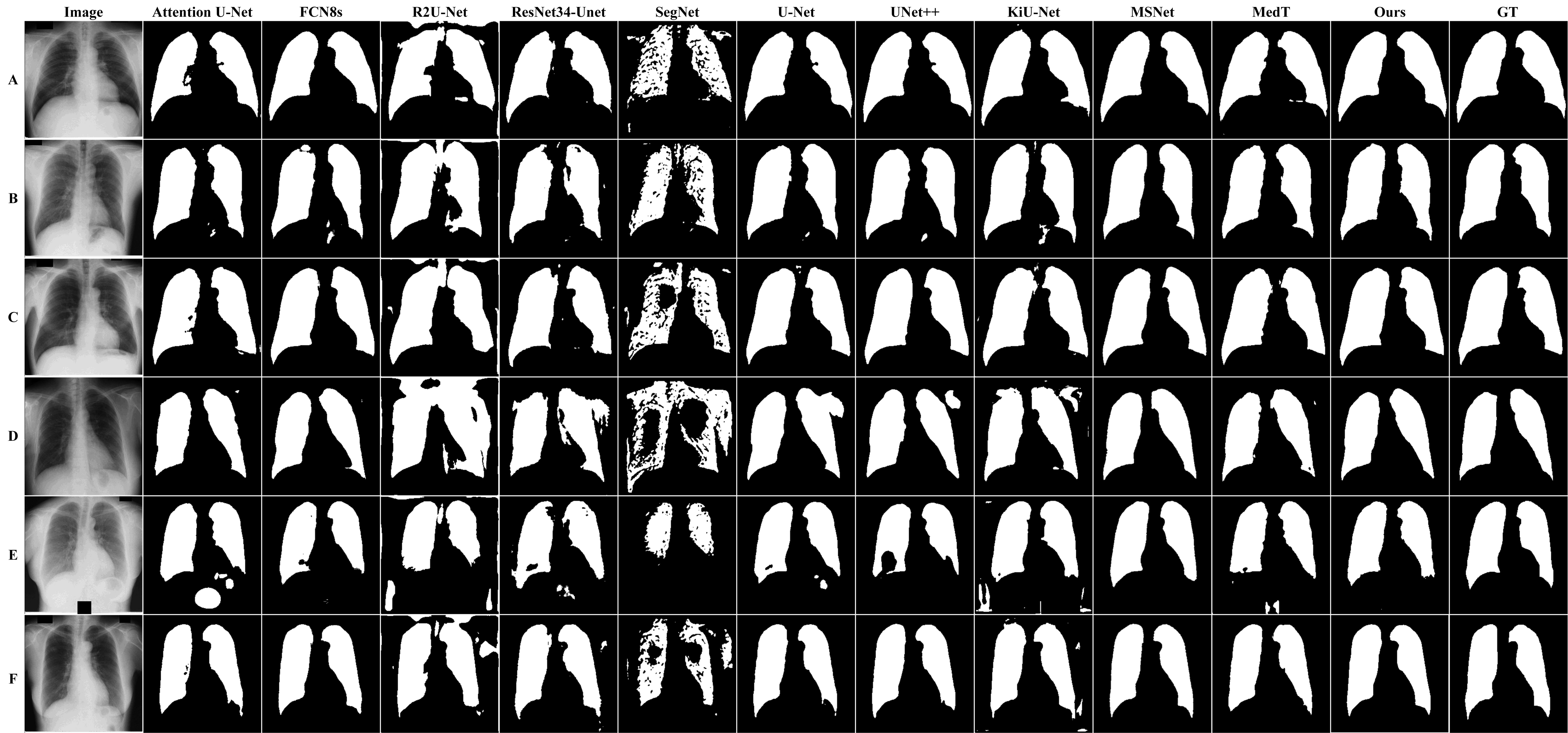}
	\caption{In the comparative experiment, the prediction results of each network on the JSRT dataset.}
	\label{fig9}
\end{figure*}

As in Fig.~\ref{fig9}, we show the prediction results of each method on the JSRT dataset. As in sample A, there is interference between the central positions of the left and right lungs due to obscuration by the heart. Several methods such as Attention U-Net\cite{39}, U-Net, and UNet++ showed weakness in processing the A samples and were unable to correlate the features in the occluded part with those in other locations, so the prediction results differed significantly from GT. In contrast Ours showed a better ability. The results from the processing of the B and E samples show that the quality of the prediction maps obtained by networks such as KiU-Net, ResNet34-Unet\cite{40}, and R2U-Net\cite{41} is severely disturbed due to interference from other organs as well as environmental factors. In comparison to these methods, Ours shows good performance, with the module created and used by Ours effectively suppressing the weight values of unimportant features and boosting the weight values of the target features, resulting in results similar to those of GT. Comparing the overall results of all networks longitudinally, the results achieved by MSNet are the best of all the comparison networks, but there is still a gap compared to Ours. For the D sample, MSNet clearly differed from GT for the right lung.

\begin{figure}
	\centering
	\includegraphics[scale=0.215]{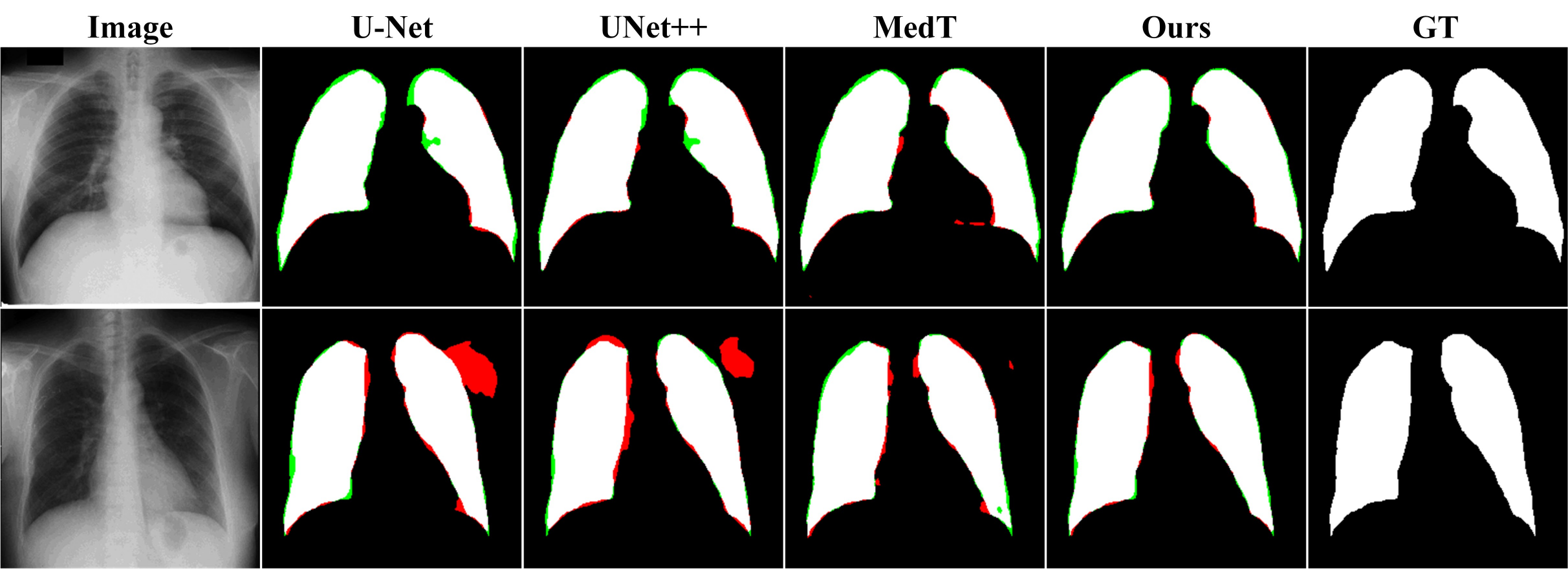}
	\caption{Visual overlap comparison chart for the JSRT dataset. Red for incorrectly identified areas, green for unidentified areas.}
	\label{fig10}
\end{figure}

To highlight the ability to process edge features, we selected samples A and D for visual comparison. As in Fig.~\ref{fig10}, we visually overlapped the predicted result maps of each method with the ground truth (GT), where the red regions represent misidentified areas and the green regions represent unidentified areas. From the comparison results, although our method still has some shortcomings in handling edge features, the achieved results are higher than those of the U-Net, UNet++, and MedT networks. From the processing of samples in Fig.~\ref{fig10}, DmADs-Net shows better ability to handle non-target features.

\begin{figure}
	\centering
	\includegraphics[scale=0.215]{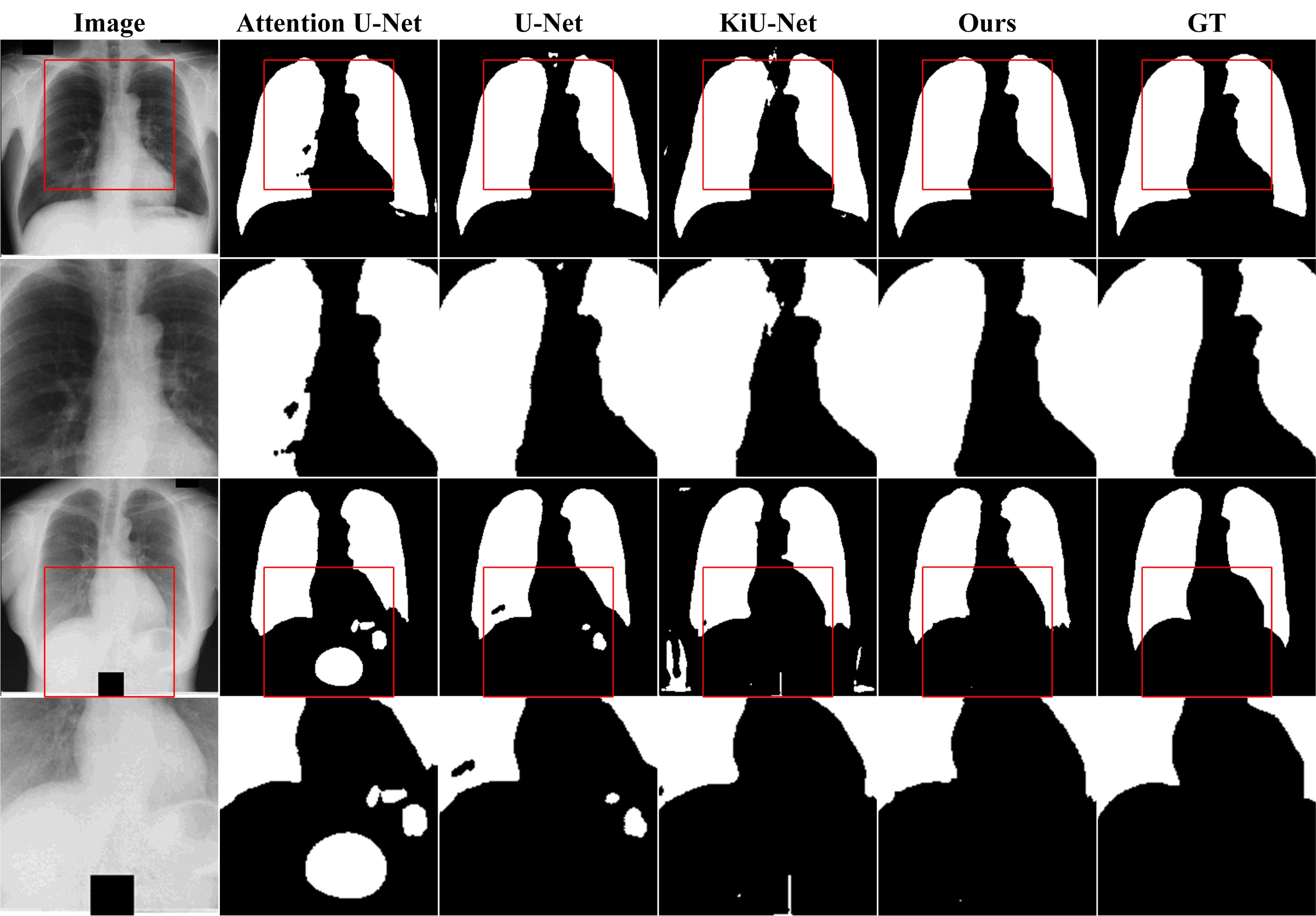}
	\caption{Local zoom comparison of the JSRT dataset.}
	\label{fig11}
\end{figure}

As in Fig.~\ref{fig11}, we select C sample for local feature amplification comparison, from the amplification results, KiU-Net, Attention U-Net for the processing of edge features obviously has shortcomings. Compared to the first two networks, U-Net achieves significantly better processing results on edge features, but is still less refined than Ours.

\begin{table}
	\renewcommand\arraystretch{1.5}
	\caption{Metrics for each network in the JSRT dataset}
	\setlength{\tabcolsep}{2.5pt}
	\begin{center}
		\begin{tabular}{lcccc}
			\hline
			\multicolumn{1}{c}{\textbf{Method}} & \textbf{Dice} & \textbf{IoU} & \textbf{Precision} & \textbf{Recall} \\
			\hline
			\textbf{Attention U-Net\cite{39}} & 96.17 & 92.65 & {\color[HTML]{FF0000} \textbf{98.27}} & 94.23 \\
			\textbf{FCN8s\cite{43}} & 96.53 & 93.32 & 97.06 & 96.63 \\
			\textbf{R2U-Net\cite{41}} & 86.22 & 76.09 & 78.56 & 96.29 \\
			\textbf{ResNet34-Unet\cite{40}} & 95.21 & 90.9 & 95.92 & 94.71 \\
			\textbf{SegNet\cite{9}} & 69.23 & 54.79 & 81.9 & 60.89 \\
			\textbf{U-Net\cite{11}} & 95.68 & 91.79 & 97.12 & {\color[HTML]{0070C0} \textbf{97.21}} \\
			\textbf{UNet++\cite{12}} & 96.58 & 93.41 & 96.40 & 96.36 \\
			\textbf{KiU-Net\cite{13}} & 96.48 & 93.25 & 95.66 & 96.46 \\
			\textbf{MSNet\cite{42}} & {\color[HTML]{0070C0} \textbf{97.29}} & {\color[HTML]{0070C0} \textbf{94.23}} & {\color[HTML]{0070C0} \textbf{97.74}} & 97.13 \\
			\textbf{MedT\cite{25}} & 95.87 & 92.10 & 96.59 & 94.44 \\
			\textbf{Ours} & {\color[HTML]{FF0000} \textbf{97.66}} & {\color[HTML]{FF0000} \textbf{95.44}} & 97.59 & {\color[HTML]{FF0000} \textbf{97.96}}\\ 
			\hline
		\end{tabular}
		\label{tab:table1}
	\end{center}
\end{table}

Table 1 shows the metric results of various comparative networks on the JSRT dataset. As seen from the results, MSNet achieved the highest results among all comparative networks, which is consistent with our subjective evaluation of the prediction results. However, there is still a gap between MSNet and our proposed method. Ours outperformed the second-ranked network by 0.37 in Dice score, 1.21 points in IoU score, and 0.75 in Recall. Based on both subjective and objective evaluations, our method achieved satisfactory results on the JSRT dataset. Although there is still room for improvement in edge feature processing, our method outperforms the mainstream networks currently available.

\begin{figure*}
	\centering
	\includegraphics[scale=0.20]{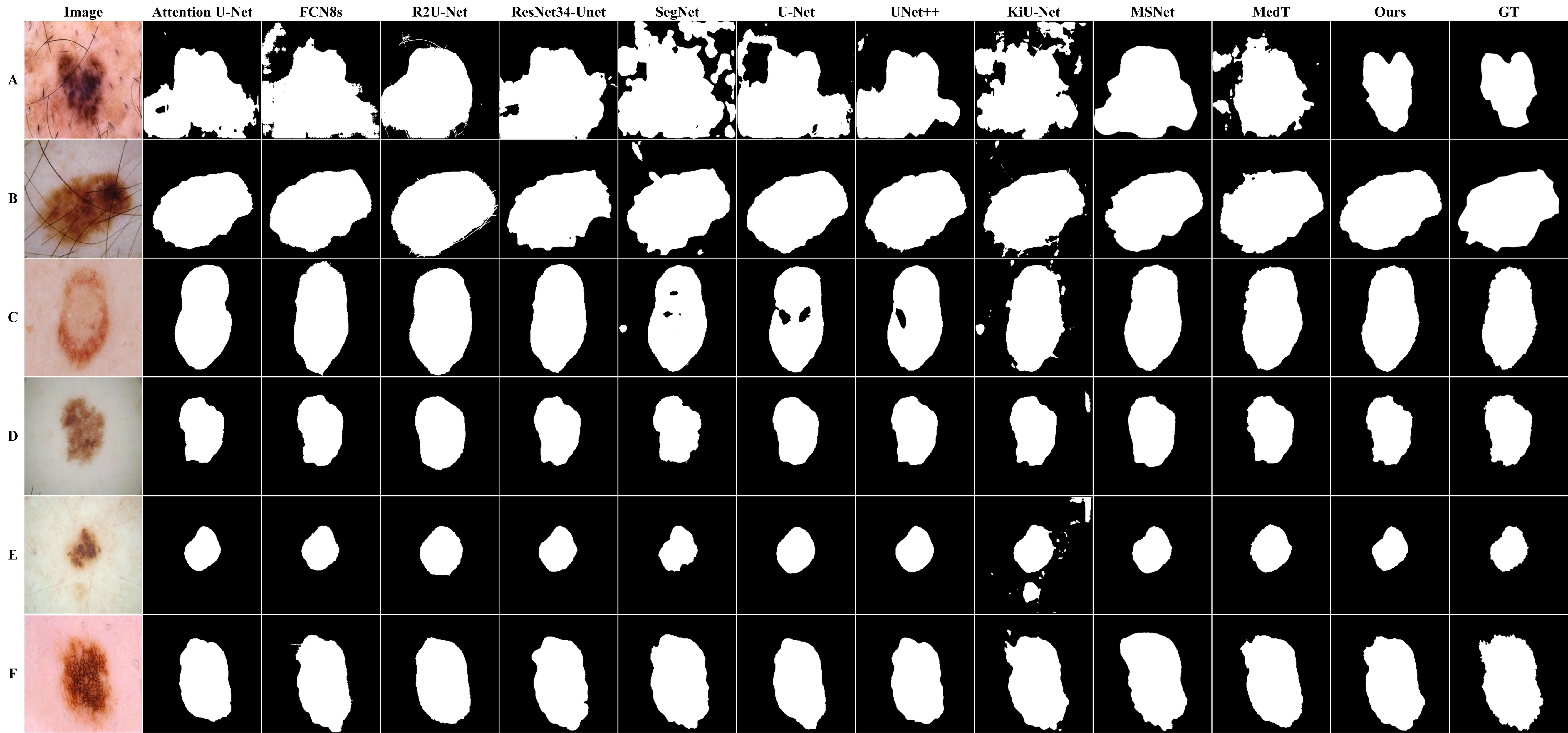}
	\caption{In the comparative experiment, the prediction results of each network on the ISIC2016 dataset.}
	\label{FIG:12}
\end{figure*}

\subsubsection{ISIC2016}

Fig.~\ref{fig12} depicts the segmentation results achieved by different methods on the ISIC2016 dataset. Given the dataset's large number of samples in different environments, we have carefully selected the most representative ones for display. Sample A and B both contain a significant amount of hair within the segmentation target, while Sample A also includes many patches around it. From the segmentation results, we observe that R2U-Net, U-Net, and other methods are impacted by hair, leading to evident errors in the target's segmentation. Similarly, KiU-Net, MedT, and other approaches are also affected by environmental factors, resulting in varying degrees of errors in boundary determination. Furthermore, the segmentation target in Sample C has highly similar texture features to the surrounding environment, and the edges are relatively blurred, demanding high network performance. SegNet and MedT exhibit obvious shortcomings in the segmentation results of Sample C. Ours exhibits relatively good performance, outperforming current state-of-the-art networks in both target localization and edge feature extraction.

\begin{figure}
	\centering
	\includegraphics[scale=0.215]{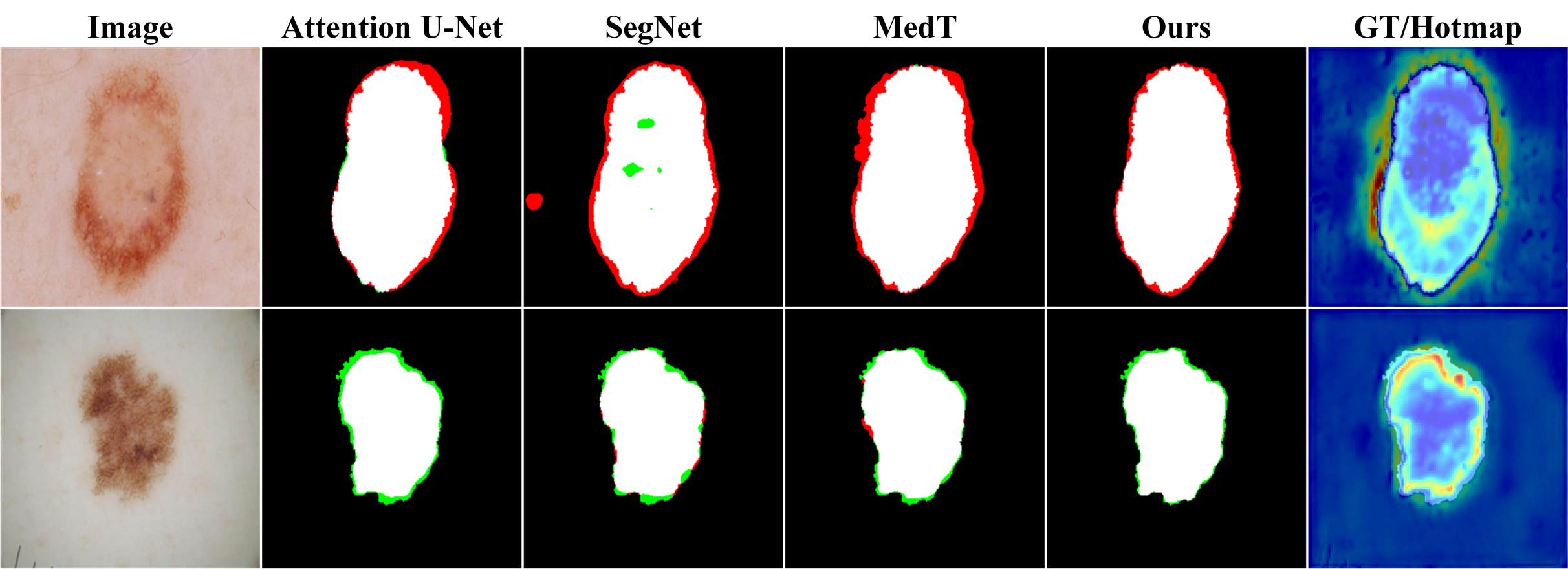}
	\caption{Visual overlap comparison chart for the ISIC2016 dataset. Red for incorrectly identified areas, green for unidentified areas.}
	\label{fig13}
\end{figure}

To highlight the ability to process edge features, we chose Attention U-Net, SegNet, MedT, and Ours for visual overlap comparisons, as in Fig.~\ref{fig13}. In the processing of sample C, the three comparison networks are significantly too coarse for the target edge features, while Ours is much closer to GT in its determination of fuzzy edges. Overlaying the feature attention plots of the final Ours results with GT, it can be seen that Ours is much more refined in its determination of edge features, and is almost identical to GT in the lower half of the sample, although there are also deviations from GT in the upper half of the D sample.

\begin{figure}
	\centering
	\includegraphics[scale=0.215]{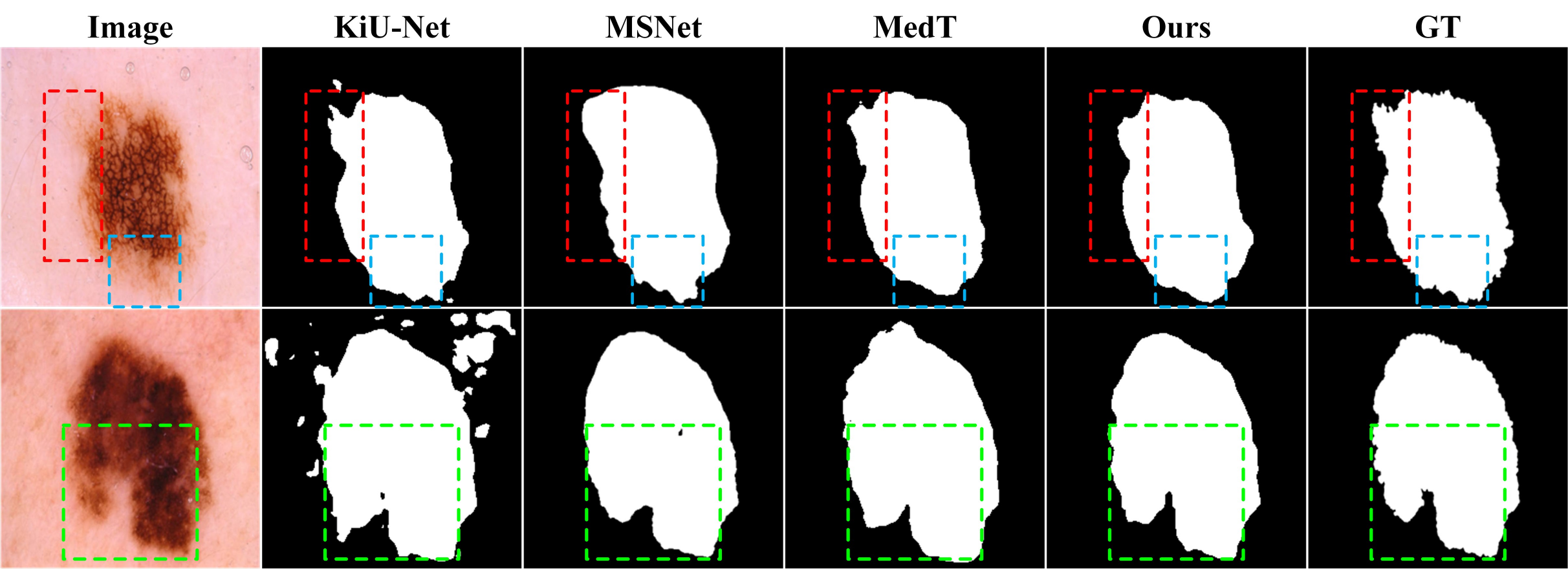}
	\caption{Comparison of local features of ISIC2016 dataset.}
	\label{fig14}
\end{figure}

As in Fig.~\ref{fig14}, we select the F sample and supplement it with a set of features with complex boundaries for local feature comparison. From the processing of the red and blue box positions in the figure by each network, MSNet is too rounded for samples with blurred edges, and the results obtained by KiU-Net and MedT are similar to those of Ours, but from the processing of the green box positions, the results obtained by Ours are most similar to those of GT.

\begin{figure}
	\centering
	\includegraphics[scale=0.83]{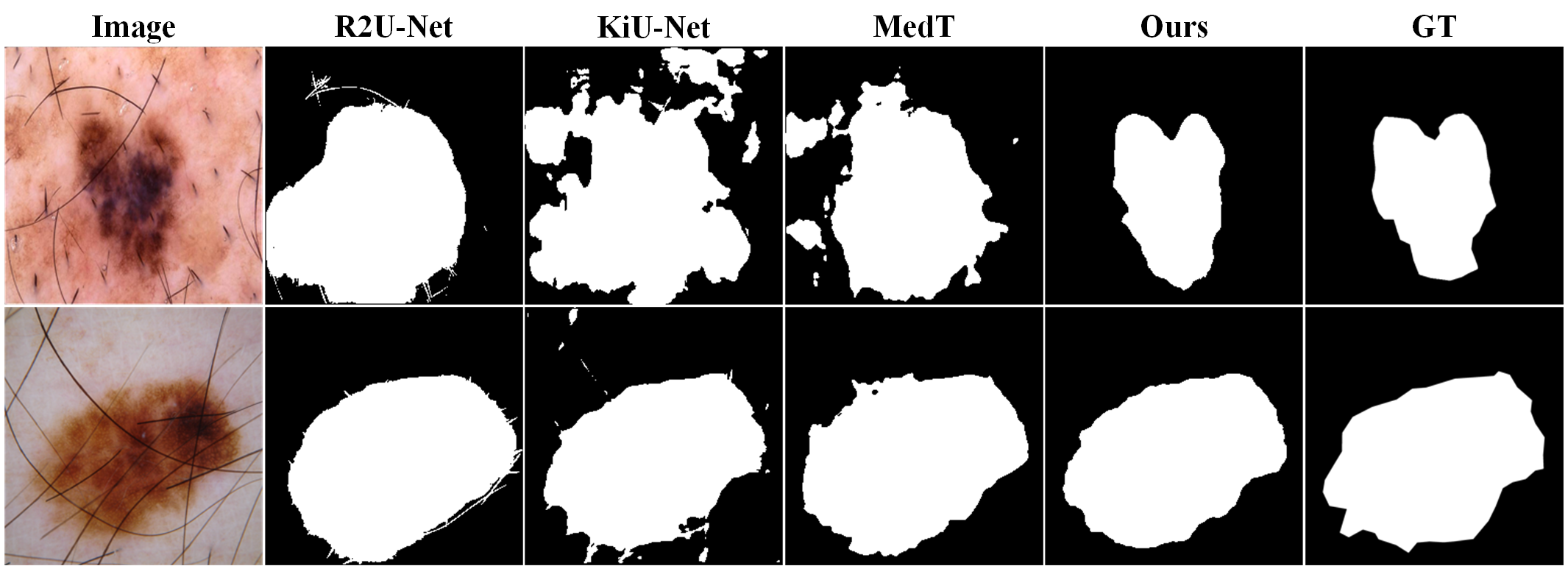}
	\caption{Comparison of environmental disturbances in the ISIC2016 dataset.}
	\label{fig28}
\end{figure}

Fig.\ref{fig28} displays two sets of samples containing environmental interference. In their processing, R2U-Net, KiU-Net, and MedT all exhibit lower generalization capabilities and poorer robustness. Although the results obtained by Ours also have some discrepancies from the GT, they do not include environmental information compared to the result images of the other three methods. This indicates that the model effectively distinguishes between the target and the environment in the process of enhancing detail information, thanks to the multi-scale receptive field design in MSCFA. Moreover, the enhancement of macro-location information of the target in LFA promotes the model's perceptual ability towards the target. The information enhancement and fusion by FRFB ultimately achieve the reconstruction of the target. Eventually, Ours achieves results closest to the GT.


\begin{table}
	\renewcommand\arraystretch{1.5}
	\caption{Metrics for each network in the ISIC2016 dataset}
	\setlength{\tabcolsep}{2.5pt}
	\begin{center}
		\begin{tabular}{lcccc}
			\hline
			\multicolumn{1}{c}{\textbf{Method}} & \textbf{Dice} & \textbf{IoU} & \textbf{Precision} & \textbf{Recall} \\
			\hline
			\textbf{Attention U-Net\cite{39}} & 88.34 & 81.22 & {\color[HTML]{FF0000} \textbf{93.77}} & 87.32 \\
			\textbf{FCN8s\cite{43}} & 90.68 & 84.61 & 90.28 & {\color[HTML]{FF0000} \textbf{94.20}} \\
			\textbf{R2U-Net\cite{41}} & 87.60 & 80.25 & 88.00 & 92.07 \\
			\textbf{ResNet34-Unet\cite{40}} & {\color[HTML]{0070C0} \textbf{90.74}} & {\color[HTML]{0070C0} \textbf{84.75}} & 91.20 & {\color[HTML]{0070C0} \textbf{93.44}} \\
			\textbf{SegNet\cite{9}} & 76.38 & 66.57 & 79.15 & 85.71 \\
			\textbf{U-Net\cite{11}} & 88.66 & 81.92 & 90.48 & 91.17 \\
			\textbf{UNet++\cite{12}} & 90.20 & 83.83 & 92.94 & 90.68 \\
			\textbf{KiU-Net\cite{13}} & 83.22 & 75.03 & 81.83 & 91.80 \\
			\textbf{MSNet\cite{42}} & 89.40 & 83.22 & 91.54 & 91.52 \\
			\textbf{MedT\cite{25}} & 88.95 & 82.03 & 90.11 & 91.78 \\
			\textbf{Ours} & {\color[HTML]{FF0000} \textbf{92.64}} & {\color[HTML]{FF0000} \textbf{86.3}} & {\color[HTML]{0070C0} \textbf{93.34}} & 93.00 \\
			\hline
		\end{tabular}
		\label{tab:table2}
	\end{center}
\end{table}

Table 2 shows the metric results of various methods tested on the ISIC2016 dataset. Ours achieved first place in Dice and IoU-score, and second place in Precision. Combined with subjective evaluation, Ours indeed achieved satisfactory results on the ISIC2016 dataset. Despite facing complex environments and diverse samples, Ours still exhibited strong generalization ability. In addition, through comparative experiments, we also found that there is still room for improvement in Ours in terms of feature extraction and boundary attention.

\subsubsection{DSB2018}

\begin{figure*}
	\centering
	\includegraphics[scale=0.20]{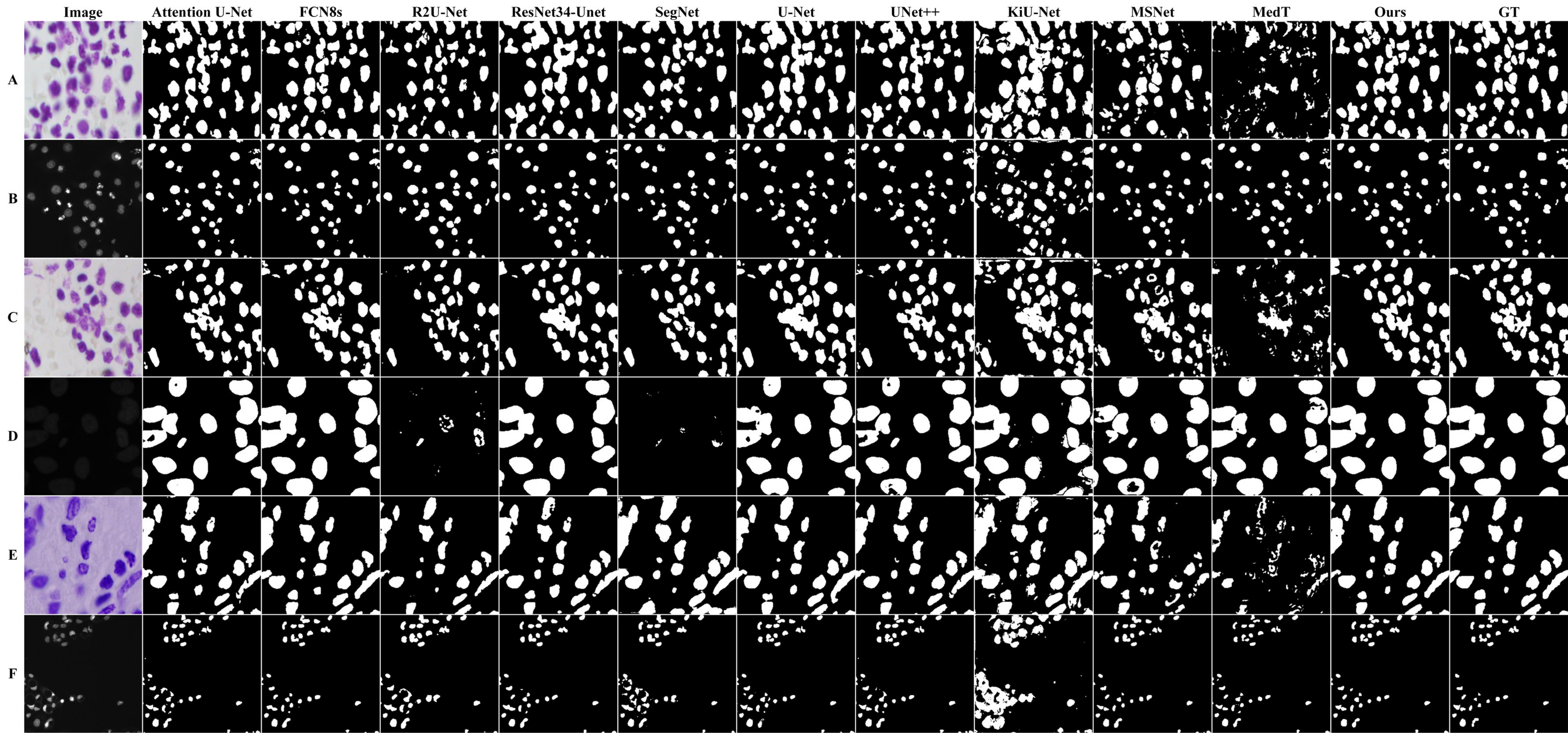}
	\caption{In the comparative experiment, the prediction results of each network on the DSB2018 dataset.}
	\label{fig15}
\end{figure*}

The DSB2018 dataset contains a large number of samples in different environments, and therefore, we selected some results for display, as shown in Fig.~\ref{fig15}. Although samples A and C are stained with H$\&$E, the boundaries of some cells are still not clear, as evidenced by the R2U-Net results where the volume of individual cells is significantly smaller than that in the ground truth (GT). In contrast, although U-Net's segmentation results show little difference in cell volume compared to GT, U-Net cannot identify the edge features of individual cells in dense cell clusters, resulting in severe cell adhesion in the processing results of samples A and C. In the processing of sample E, ResNet34-Unet clearly lacks an ambiguously stained cell, while KiU-Net's handling of cell boundaries is too rough. We use a longitudinal comparison to analyze the stability of each method, where MedT shows relatively poor stability, and its handling of unstained samples is significantly higher than that of stained samples. R2U-Net and SegNet's feature extraction for the unstained sample D is significantly lower than that of all the networks involved in comparison.

\begin{figure}
	\centering
	\includegraphics[scale=0.215]{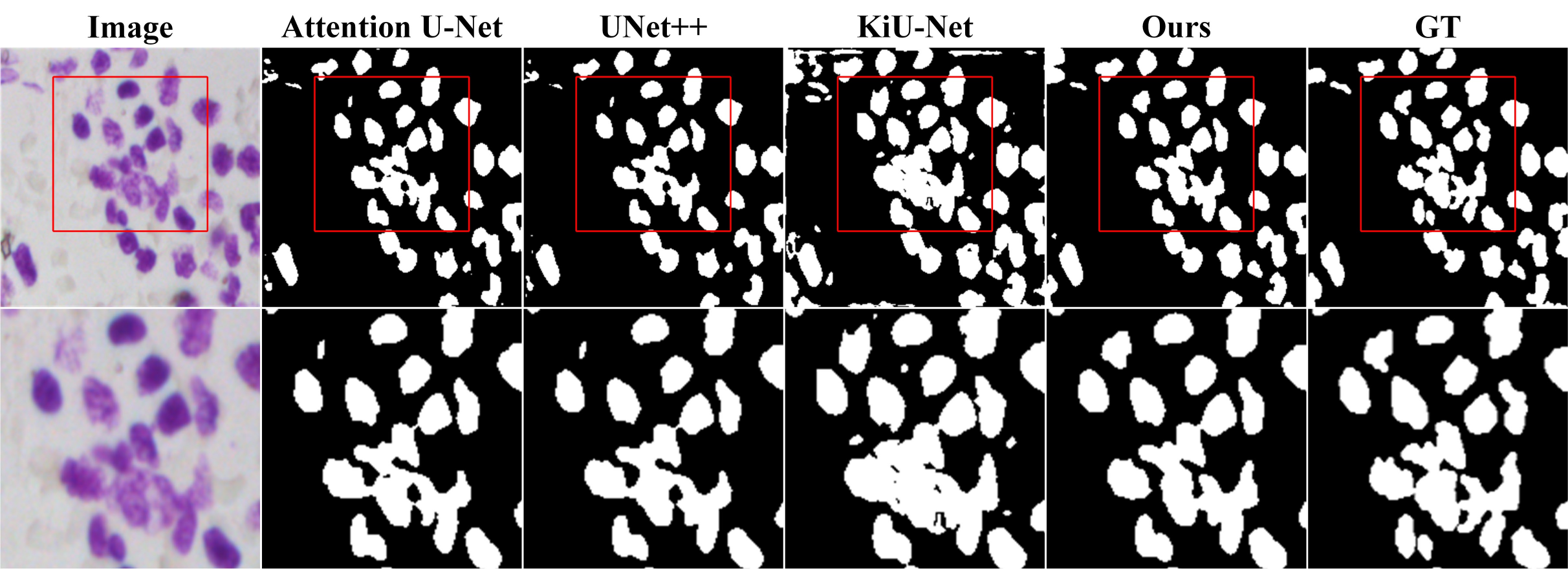}
	\caption{Local zoom comparison of the DSB2018 dataset.}
	\label{fig16}
\end{figure}

As in Fig.~\ref{fig16}, Sample C is selected for local comparison in order to highlight the processing results of the local features. We zoom in on the locations shown in red boxes, and the processing of local features shows that Ours is significantly more capable of focusing on weak features than the networks involved in the comparison. Attention U-Net and UNet++ do not complete the segmentation of lightly stained cells, while our network is able to identify cells with weak features more completely.

\begin{figure}
	\centering
	\includegraphics[scale=0.215]{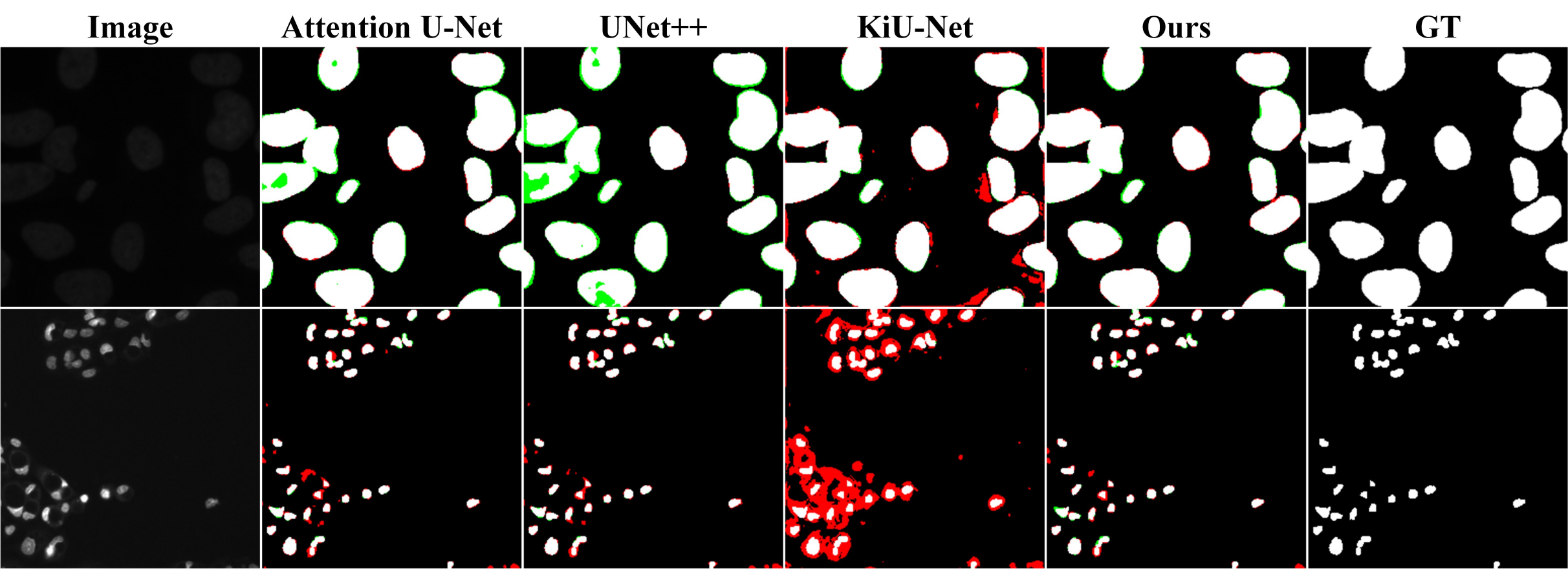}
	\caption{Visual overlap comparison chart for the DSB2018 dataset. Red for incorrectly identified areas, green for unidentified areas.}
	\label{fig17}
\end{figure}

As in Fig.~\ref{fig17}, we select two sets of samples for visual overlap comparison, and the overall effect shows that each network is able to complete segmentation of the cells. The cells in the first sample are larger in size, while in the second sample they are smaller and more densely arranged. KiU-Net handles the second sample significantly worse, in contrast to our network which shows better performance both for segmentation of large targets and for segmentation of small and dense cells.

\begin{table}
	\renewcommand\arraystretch{1.5}
	\caption{Metrics for each network in the DSB2018 dataset}
	\setlength{\tabcolsep}{2.5pt}
	\begin{center}
		\begin{tabular}{lcccc}
			\hline
			\multicolumn{1}{c}{\textbf{Method}} & \textbf{Dice} & \textbf{IoU} & \textbf{Precision} & \textbf{Recall} \\
			\hline
			\textbf{Attention U-Net\cite{39}} & 88.58 & 81.22 & 89.30 & 89.15 \\
			\textbf{FCN8s\cite{43}} & 89.23 & {\color[HTML]{0070C0} \textbf{82.18}} & 90.38 & {\color[HTML]{0070C0} \textbf{90.66}} \\
			\textbf{R2U-Net\cite{41}} & 79.48 & 68.94 & 89.39 & 78.07 \\
			\textbf{ResNet34-Unet\cite{40}} & {\color[HTML]{0070C0} \textbf{90.14}} & {\color[HTML]{0070C0} \textbf{82.18}} & 91.15 & 88.49 \\
			\textbf{SegNet\cite{9}} & 77.27 & 66.92 & 92.37 & 72.16 \\
			\textbf{U-Net\cite{11}} & 88.70 & 80.45 & 90.11 & 85.88 \\
			\textbf{UNet++\cite{12}} & 89.46 & 81.85 & 90.31 & 88.46 \\
			\textbf{KiU-Net\cite{13}} & 87.68 & 78.56 & 89.25 & 85.19 \\
			\textbf{MSNet\cite{42}} & 85.01 & 75.56 & 89.81 & 82.32 \\
			\textbf{MedT\cite{25}} & 80.59 & 71.41 & {\color[HTML]{FF0000} \textbf{92.91}} & 73.62 \\
			\textbf{Ours} & {\color[HTML]{FF0000} \textbf{91.42}} & {\color[HTML]{FF0000} \textbf{84.90}} & {\color[HTML]{0070C0} \textbf{92.90}} & {\color[HTML]{FF0000} \textbf{91.05}} \\
			\hline
		\end{tabular}
		\label{tab:table3}
	\end{center}
\end{table}

Table 3 shows the scores of all networks compared in the experiment, and our network achieved the highest scores in the Dice, IoU-score, and Recall metrics. It outperformed the second place by 1.28 in Dice and 2.72 in IoU-score. Combined with subjective analysis on the DSB2018 dataset, DmADs-Net demonstrated excellent segmentation performance in complex environments, with superior ability to handle local features compared to other methods. It is worth noting that from the visual comparison and local feature enlargement analysis, there is still room for improvement in DmADs-Net's edge feature extraction and target localization. This will be the focus of our future research.

\subsubsection{BUSI}

\begin{figure*}
	\centering
	\includegraphics[scale=0.20]{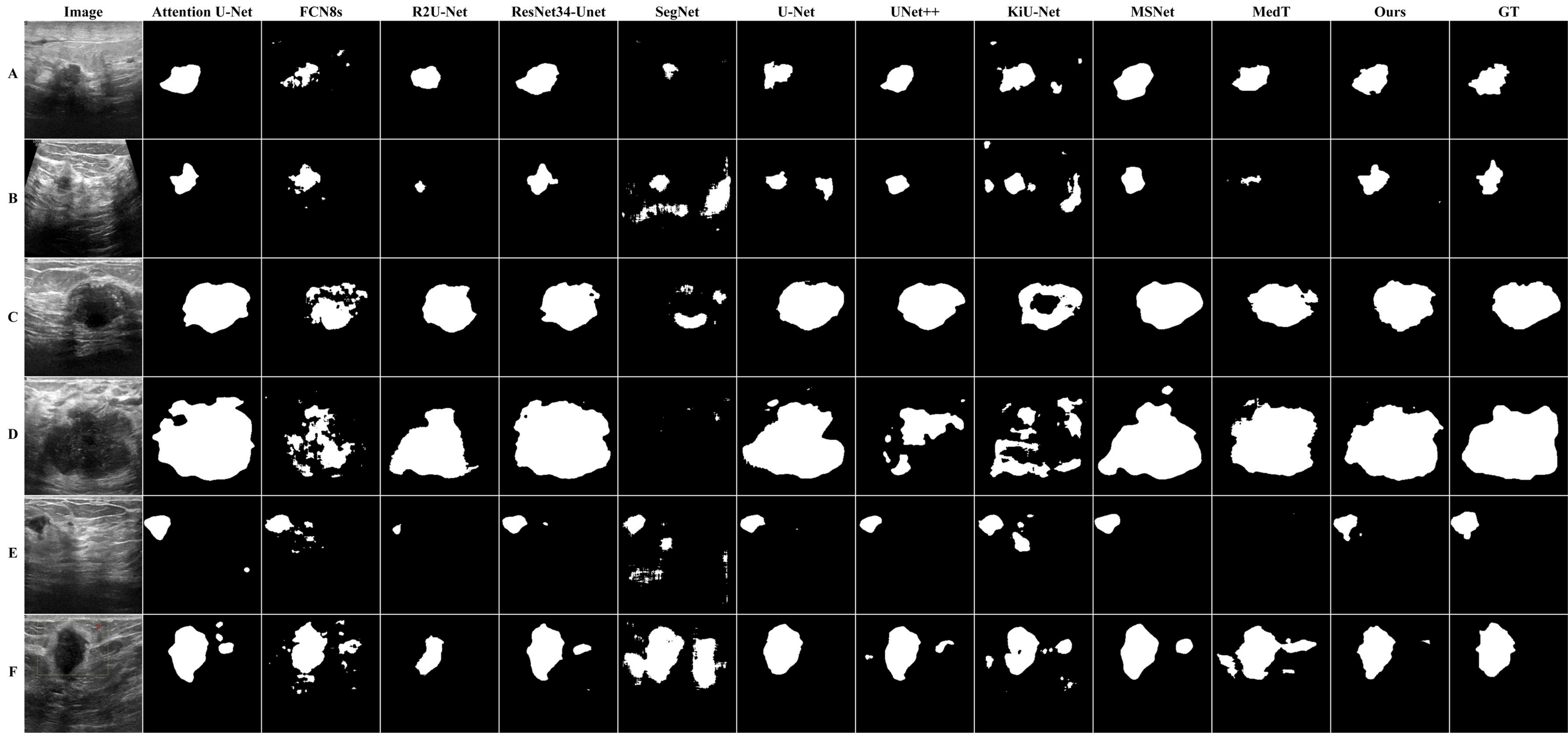}
	\caption{In the comparative experiment, the prediction results of each network on the BUSI dataset.}
	\label{fig18}
\end{figure*}

Fig.~\ref{fig18} shows the results of each method for the malignant breast cancer samples. Compared to the first three datasets, the samples in the BUSI dataset have highly similar textural features to their surroundings and the segmentation targets in all samples are characterised by blurred edge features. From the processing of the C and D samples, networks such as UNet++, MSNet and MedT are able to locate the position of the segmentation target, but do not achieve satisfactory results in edge determination. Again, Ours suffers from the same problem, but in contrast, the predictions obtained by Ours are most similar to those of GT.

\begin{figure}
	\centering
	\includegraphics[scale=0.215]{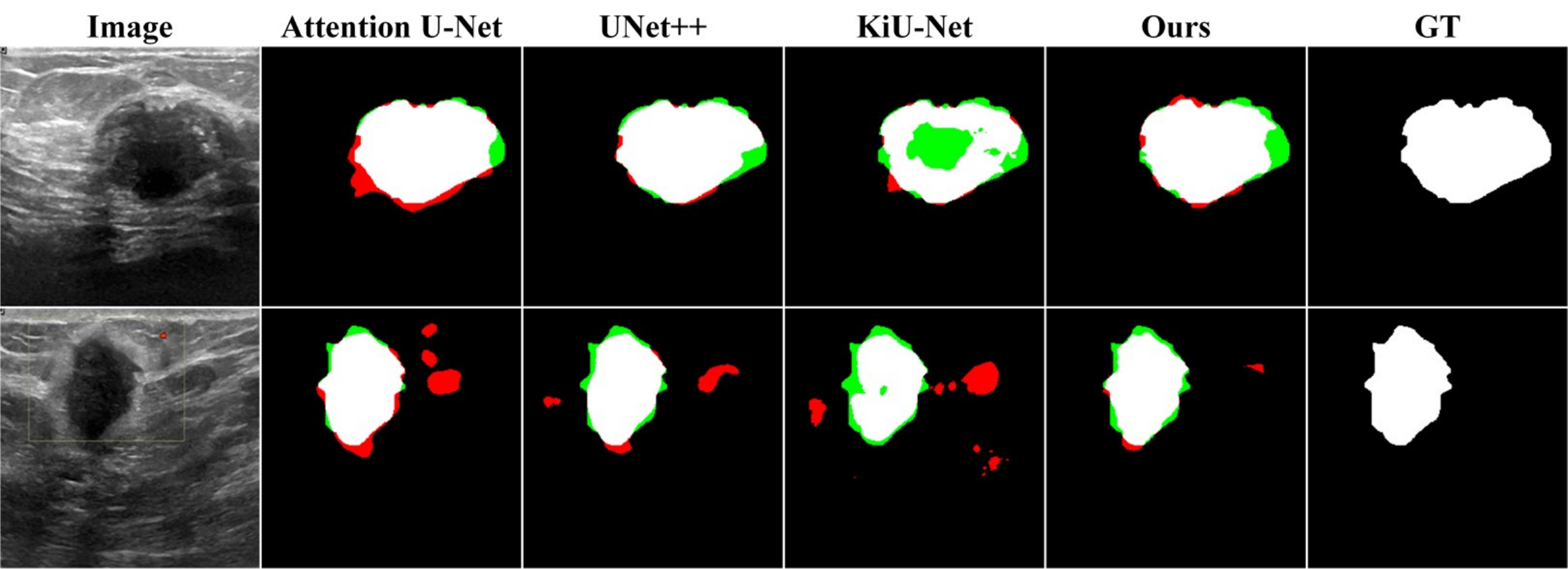}
	\caption{Visual overlap comparison chart for the BUSI dataset. Red for incorrectly identified areas, green for unidentified areas.}
	\label{fig19}
\end{figure}

In order to highlight the localization and edge processing capabilities of each network for the predicted targets, we selected C and F samples for visual overlap processing, as in Fig.~\ref{fig19}. In the first sample, our network and UNet++ handle the edges in the lower left corner of the target relatively well, while for the right side, all four networks show varying degrees of missing features.

\begin{table}
	\renewcommand\arraystretch{1.5}
	\caption{Metrics for each network in the BUSI dataset}
	\setlength{\tabcolsep}{2.5pt}
	\begin{center}
		\begin{tabular}{lcccc}
				\hline
				\multicolumn{1}{c}{\textbf{Method}} & \textbf{Dice} & \textbf{IoU} & \textbf{Precision} & \textbf{Recall} \\
				\hline
				\textbf{Attention U-Net\cite{39}} & 61.54 & 50.13 & 68.12 & 64.64 \\
				\textbf{FCN8s\cite{43}} & 54.04 & 40.3 & 68.15 & 50.75 \\
				\textbf{R2U-Net\cite{41}} & 31.87 & 22.97 & 63.64 & 24.44 \\
				\textbf{ResNet34-Unet\cite{40}} & {\color[HTML]{0070C0} \textbf{66.39}} & 54.21 & {\color[HTML]{0070C0} \textbf{72.16}} & 68.19 \\
				\textbf{SegNet\cite{9}} & 30.29 & 20.85 & 58.14 & 29.06 \\
				\textbf{U-Net\cite{11}} & 61.73 & 48.76 & 68.48 & 62.40 \\
				\textbf{UNet++\cite{12}} & 61.29 & 49.67 & {\color[HTML]{FF0000} \textbf{74.14}} & 58.66 \\
				\textbf{KiU-Net\cite{13}} & 49.98 & 35.91 & 62.22 & 48.38 \\
				\textbf{MSNet\cite{42}} & 66.10 & {\color[HTML]{0070C0} \textbf{54.85}} & 66.31 & {\color[HTML]{FF0000} \textbf{74.40}} \\
				\textbf{MedT\cite{25}} & 55.81 & 44.54 & 67.43 & 55.23 \\
				\textbf{Ours} & {\color[HTML]{FF0000} \textbf{71.35}} & {\color[HTML]{FF0000} \textbf{55.46}} & 71.14 & {\color[HTML]{0070C0} \textbf{68.46}}\\
				\hline
			\end{tabular}
			\label{tab:table4}
		\end{center}
\end{table}

As in Table 4, we show the experimental metrics on the BUSI dataset. Feature extraction is difficult due to the presence of the same texture features in the breast cancer data samples as in the surrounding environment. Ours still achieves the highest scores on Dice, IoU-score. Combined with the subjective analysis, Ours' experiments on the BUSI dataset achieved good results, but still need to explore more in edge feature processing as well as global feature information to obtain even better results.

\subsubsection{GlaS}

\begin{figure*}
	\centering
	\includegraphics[scale=0.20]{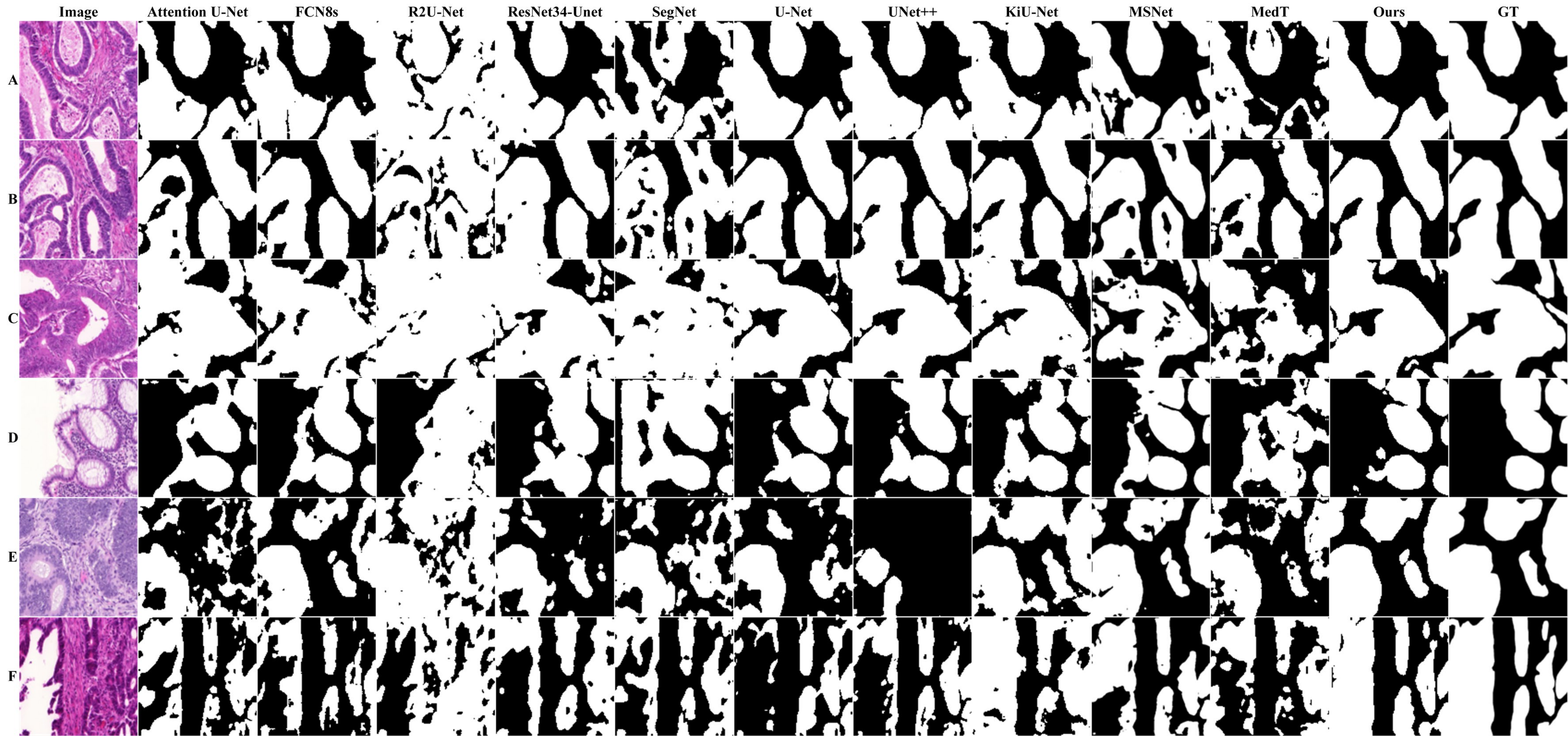}
	\caption{In the comparative experiment, the prediction results of each network on the GlaS dataset.}
	\label{fig20}
\end{figure*}

Fig.~\ref{fig20} shows a comparison experiment on the GlaS dataset, where we have selected representative samples with different shapes for presentation. Sample A has well-defined boundaries and most of the networks are more accurate for feature identification, but the processing results obtained by MedT, MSNet, and SegNet are not as good as our network for correlating intracellular regions. The segmentation target in the C sample has the same texture characteristics as its surroundings and it is difficult to discriminate the boundaries by visual observation alone, whereas the shape and boundary information of the target can be easily obtained by excellent automatic segmentation techniques. Networks such as UNet++, U-Net and KiU-Net are better for cell segmentation, while networks such as Attention U-Net and FCN8s clearly do not handle boundary feature information well. In contrast, our network is able to identify edge information more accurately, with the highest similarity to GT. Of course, Ours does also have room for improvement, for example, the small cells on the right side of the GT plot in the C sample, which Ours does not identify. There is a large difference between the left and right environments in the D sample, with cells concentrated on the right side of the sample, while the left environment produces the same boundary information as the cells due to cell breakage, which places a high demand on the network's feature recognition capability. All networks are affected by the left-hand side of the environment for cell partitioning, with Ours being the least affected.

\begin{figure}
	\centering
	\includegraphics[scale=0.215]{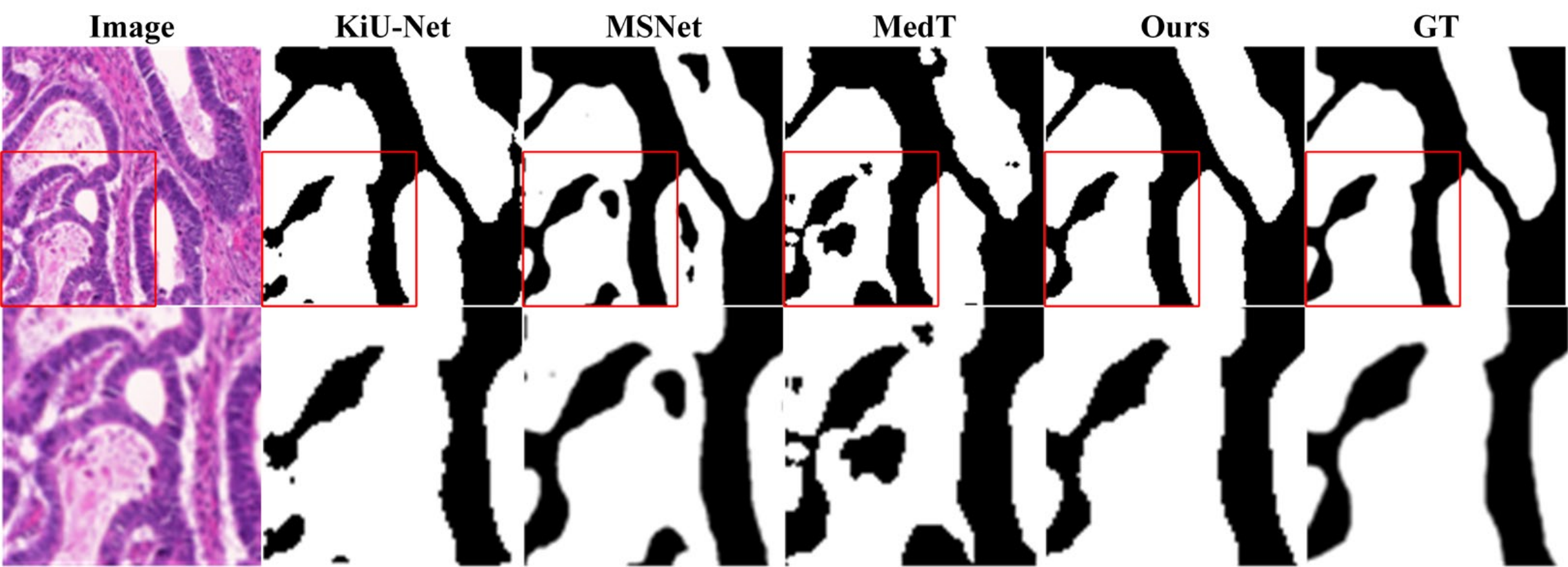}
	\caption{Local zoom comparison of the GlaS dataset.}
	\label{fig21}
\end{figure}

As in Fig.~\ref{fig21}, we select the sample B for local zooming. MSNet still has room for improvement in feature recognition, and KiU-Neth and MedT do not achieve the same results for edges as Ours. Although Ours does not achieve the same level of edge refinement as GT, it completes segmentation for all segmentation targets that appear in the sample.

\begin{figure}
	\centering
	\includegraphics[scale=0.215]{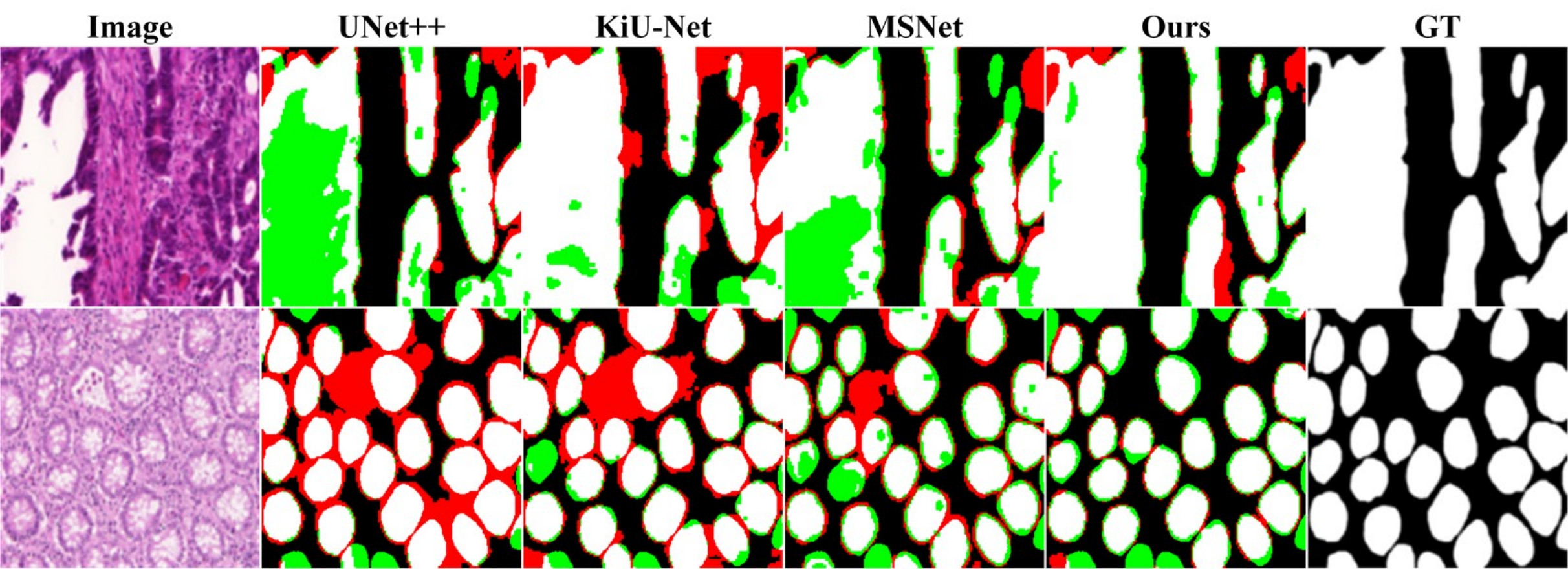}
	\caption{Visual overlap comparison chart for the GlaS dataset. Red for incorrectly identified areas, green for unidentified areas.}
	\label{fig22}
\end{figure}

Fig.~\ref{fig22} Sample F is selected and supplemented with a new sample for visual overlap comparison. The first sample has a different class of cells on the left side than on the right side and therefore presents a larger difference, which also has an impact on the segmentation results. As in UNet++, MSNet does not handle the left side cells significantly better than the right side, and although KiU-Net achieves localisation of most of the cells, there is still a gap with our network in terms of edge determination. The cells in the second sample have a more distinct boundary with their surroundings and, from the results, our network works relatively well, but there are still deficiencies in the identification of cells at the edges of the sample.

\begin{table}
	\renewcommand\arraystretch{1.5}
	\caption{Metrics for each network in the GlaS dataset}
	\setlength{\tabcolsep}{2.5pt}
	\begin{center}
		\begin{tabular}{lcccc}
			\hline
			\multicolumn{1}{c}{\textbf{Method}} & \textbf{Dice} & \textbf{IoU} & \textbf{Precision} & \textbf{Recall} \\
			\hline
			\textbf{Attention U-Net\cite{39}} & 82.74 & 71.93 & 81.34 & 84.3 \\
			\textbf{FCN8s\cite{43}} & {\color[HTML]{0070C0} \textbf{84.75}} & {\color[HTML]{0070C0} \textbf{74.41}} & {\color[HTML]{0070C0} \textbf{86.02}} & 85.29 \\
			\textbf{R2U-Net\cite{41}} & 69.99 & 55.03 & 59.24 & {\color[HTML]{FF0000} \textbf{90.92}} \\
			\textbf{ResNet34-Unet\cite{40}} & 79.39 & 67.16 & 77.45 & 84.66 \\
			\textbf{SegNet\cite{9}} & 79.84 & 67.56 & 76.18 & 86.83 \\
			\textbf{U-Net\cite{11}} & 82.62 & 71.64 & 80.42 & 83.27 \\
			\textbf{UNet++\cite{12}} & 81.47 & 71.54 & 84.08 & 84.5 \\
			\textbf{KiU-Net\cite{13}} & 82.69 & 71.7 & 80.5 & 85.22 \\
			\textbf{MSNet\cite{42}} & 83.27 & 72.61 & 85.9 & 83.39 \\
			\textbf{MedT\cite{25}} & 76.5 & 63.29 & 73.53 & 82.78 \\
			\textbf{Ours} & {\color[HTML]{FF0000} \textbf{88.21}} & {\color[HTML]{FF0000} \textbf{79.97}} & {\color[HTML]{FF0000} \textbf{90.6}} & {\color[HTML]{0070C0} \textbf{87.13}}\\
			\hline
		\end{tabular}
		\label{tab:table5}
	\end{center}
\end{table}

The metric results for each network on the GlaS dataset are shown in Table 5, with our network achieving the highest scores on Dice, IoU-score, and Precision. Ours is 3.46 points higher than the second place in the Dice, 5.56 points higher than the second place in the IoU-score and 4.58 points higher than the second place in the Precision. On the whole, Ours has shown excellent results on samples with a size of 128$\times$128, and has shown excellent results in terms of target localization and anti-environmental interference. It is also clear from the results of the comparison experiments that there is real room for improvement in our network, as in Fig.~\ref{fig22}, where the edge determination of individual cells in the second sample still needs to be enhanced.

Ours floating-point calculation amount is 33.06GMac, which is leading compared to SegNet's 457.55GMac, KiU-Net's 70.14GMac, and R2U-Net's 38.24GMac. However, compared to lightweight networks such as MedT and UNet++, Ours still needs further algorithm optimization. In terms of parameter quantity, Ours has 36.28M parameters, which is a significant advantage compared to networks such as R2U-Net and MSNet, but still insufficient compared to lightweight networks.

This concludes our comparative experiments on all five datasets, and the overall results show that Ours achieves excellent results in edge feature processing, target localisation and feature association. Although Ours outperforms current mainstream networks, the advantage is not obvious in the processing of individual samples and metrics, so we will continue our exploration of network performance.

\subsection{Ablation experiment}

This section provides an overview of the ablation experiments that will be conducted based on the network itself. The MSCFA modules will be removed from the network, and the resulting ablated model will be named DmADs-Net-a. The FRFB module will be removed and replaced with upsampling and traditional feature addition, resulting in a network named DmADs-Net-b. The LFA module will be removed and replaced with a 3x3 convolution, resulting in a network named DmADs-Net-c. All generated companion features and corresponding loss functions in the original network will be removed, resulting in a model named DmADs-Net-d. Finally, to demonstrate the performance improvement of two different ResNet structures, DmADs-Net-e will only use $ResNet_{18}$ as the backbone network. We will still conduct experiments on the five datasets used in the comparative experiments to make the ablation experiments more informative.

\subsubsection{JSRT}

\begin{figure*}
	\centering
	\includegraphics[scale=0.32]{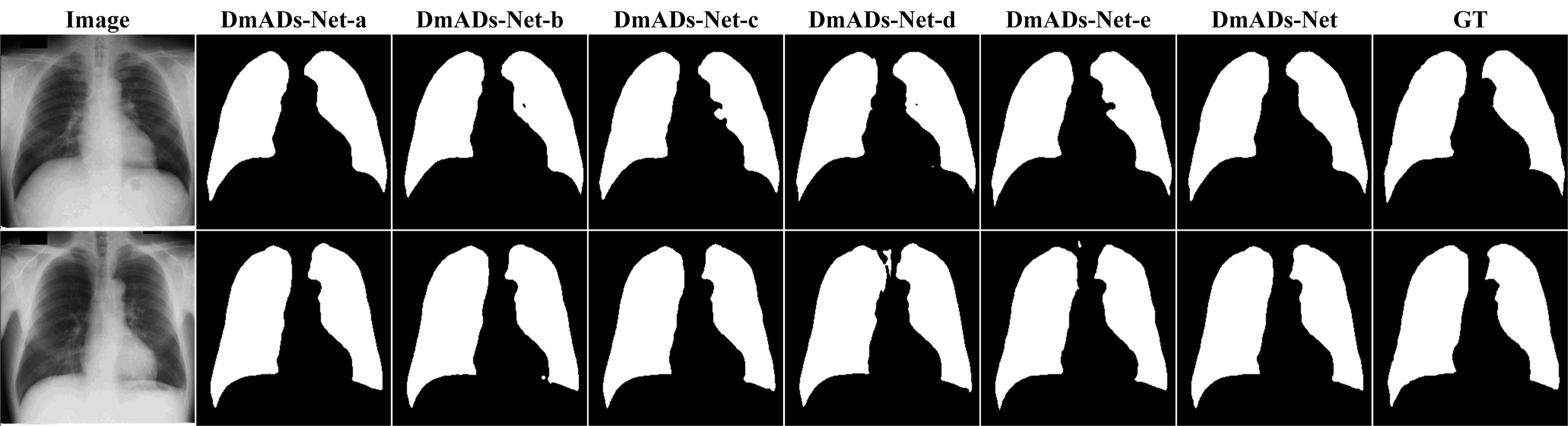}
	\caption{Results of the JSRT dataset on ablation experiment.}
	\label{fig23}
\end{figure*}

Fig.~\ref{fig23} shows the results obtained by the five ablation models and the original model on the JSRT dataset. From the analysis of the prediction results, there is a significant difference between the results obtained by DmADs-Net-c, DmADs-Net-d, and DmADs-Net-e and the GT image. Although the results obtained by the other ablation models are close to the GT, they still do not perform as well as the original network in describing edge features.

\begin{table}
	\renewcommand\arraystretch{1.5}
	\caption{Results of the JSRT dataset on ablation experiment}
	\setlength{\tabcolsep}{5.7pt}
	\begin{center}
		\begin{tabular}{lcccc}
			\hline
			\multicolumn{1}{c}{\textbf{Method}} & \textbf{Dice} & \textbf{IoU} & \textbf{Precision} & \textbf{Recall} \\
			\hline
			\textbf{DmADs-Net-a} & 97.43 & 95.20 & 97.38 & 97.86 \\
			\textbf{DmADs-Net-b} & 97.48 & 95.10 & 96.97 & 98.05 \\
			\textbf{DmADs-Net-c} & 97.37 & 94.89 & 97.98 & 96.82 \\
			\textbf{DmADs-Net-d} & 96.80 & 93.82 & 95.36 & 98.37 \\
			\textbf{DmADs-Net-e} & 97.44 & 95.02 & 95.50 & 97.43 \\
			\textbf{DmADs-Net} & 97.66 & 95.44 & 97.59 & 97.96 \\
			\hline
		\end{tabular}
		\label{tab:table6}
	\end{center}
\end{table}

As in Table 6, the results of the ablation experiments are presented. Overall, there are some differences between the indicators of the five ablation models and the original network. Combining the results of various indicators, the effectiveness of the network model is demonstrated.

\subsubsection{ISIC2016}

\begin{figure*}
	\centering
	\includegraphics[scale=0.32]{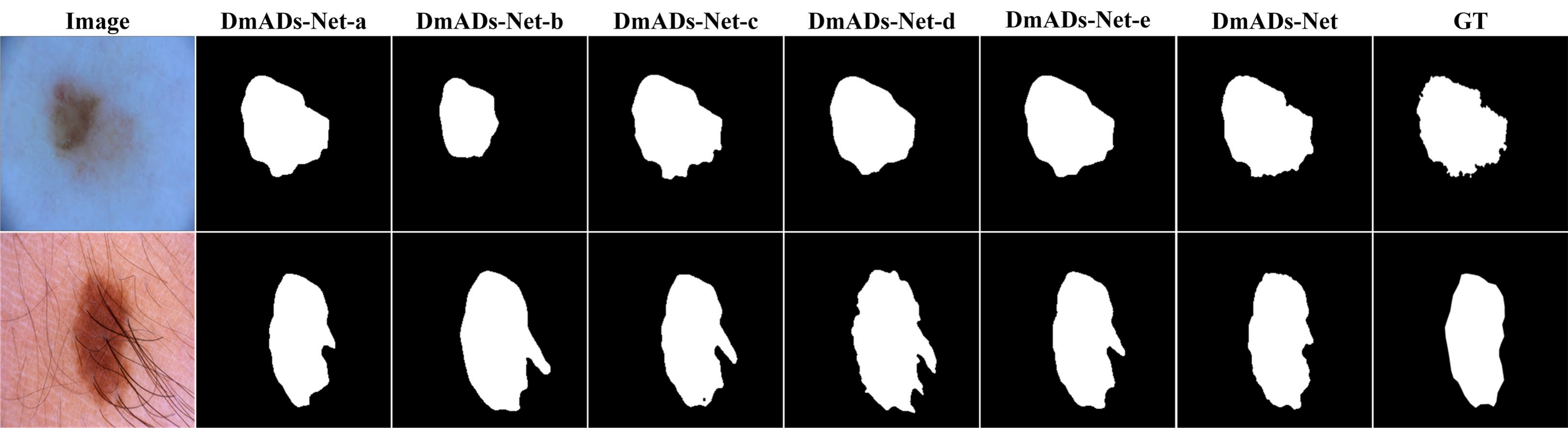}
	\caption{Results of the ISIC2016 dataset on ablation experiment.}
	\label{fig24}
\end{figure*}

As in Fig.~\ref{fig24}, we also select a representative set of samples from the ISIC2016 dataset to highlight the performance of each ablation model. For the treatment of first sample, the results obtained by DmADs-Net-b are significantly smaller. A longitudinal comparison of DmADs-Net-b shows that its ability to capture weak features in the sample is significantly lower than several other models, and that it is too rounded in its treatment of edge features, with specific edge features missing. Overall, the results obtained by the five ablation models have a certain gap with the original network.

\begin{table}
	\renewcommand\arraystretch{1.5}
	\caption{Results of the ISIC2016 dataset on ablation experiment}
	\setlength{\tabcolsep}{5.7pt}
	\begin{center}
		\begin{tabular}{lcccc}
			\hline
			\multicolumn{1}{c}{\textbf{Method}} & \textbf{Dice} & \textbf{IoU} & \textbf{Precision} & \textbf{Recall} \\
			\hline
			\textbf{DmADs-Net-a} & 91.46 & 85.39 & 94.02 & 91.27 \\
			\textbf{DmADs-Net-b} & 90.15 & 83.41 & 92.94 & 90.34 \\
			\textbf{DmADs-Net-c} & 90.57 & 84.27 & 94.19 & 89.93 \\
			\textbf{DmADs-Net-d} & 91.15 & 84.90 & 94.41 & 90.37 \\
			\textbf{DmADs-Net-e} & 91.20 & 85.01 & 92.4 & 92.54 \\
			\textbf{DmADs-Net} & 92.64 & 86.30 & 92.59 & 93.00 \\
			\hline
		\end{tabular}
		\label{tab:table7}
	\end{center}
\end{table}

As in Table 7, the metric results obtained by DmADs-Net-b and DmADs-Net-c have the largest gap with the original network. Combined with the experimental results, the results obtained by the two are also far behind the GT. Ablation experiments performed on the ISIC2016 dataset demonstrate the effectiveness of our created model.

\subsubsection{DSB2018}

\begin{figure*}
	\centering
	\includegraphics[scale=0.32]{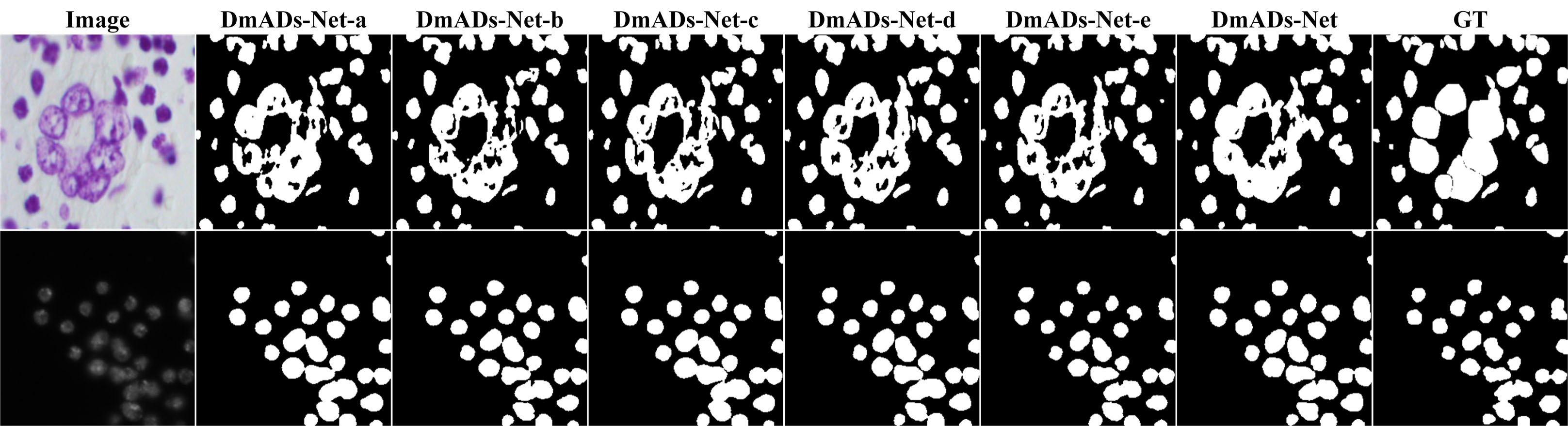}
	\caption{Results of the DSB2018 dataset on ablation experiment.}
	\label{FIG:25}
\end{figure*}

As in Fig.~\ref{FIG:25}, we have selected two sets of stained samples and two sets of unstained samples from the DSB2018 dataset for demonstration. From the treatment of the first sample, five ablation experiments showed varying degrees of feature identification problems. In combination with the second sample, the segmentation results obtained by DmADs-Net-a are more rounded, missing the edge feature information, thus allowing to see the contribution of MSCFA to the network's weak feature focus capability.

\begin{table}
	\renewcommand\arraystretch{1.5}
	\caption{Results of the DSB2018 dataset on ablation experiment}
	\setlength{\tabcolsep}{5.7pt}
	\begin{center}
		\begin{tabular}{lcccc}
			\hline
			\multicolumn{1}{c}{\textbf{Method}} & \textbf{Dice} & \textbf{IoU} & \textbf{Precision} & \textbf{Recall} \\
			\hline
			\textbf{DmADs-Net-a} & 91.31 & 84.62 & 93.77 & 89.92 \\
			\textbf{DmADs-Net-b} & 91.10 & 84.33 & 92.93 & 90.39 \\
			\textbf{DmADs-Net-c} & 91.28 & 84.43 & 91.96 & 91.75 \\
			\textbf{DmADs-Net-d} & 90.79 & 83.88 & 92.54 & 90.29 \\
			\textbf{DmADs-Net-e} & 91.24 & 84.66 & 92.91 & 90.78 \\
			\textbf{DmADs-Net} & 91.42 & 84.90 & 92.90 & 91.05 \\
			\hline
		\end{tabular}
		\label{tab:table8}
	\end{center}
\end{table}

As in Table 8, we show the metric scores obtained by the five ablation models with the original network. Combining the prediction results overall, all five ablation models obtained inferior results to the original network, which proves the effectiveness of the network for the architecture.

\subsubsection{BUSI}

\begin{figure*}
	\centering
	\includegraphics[scale=0.32]{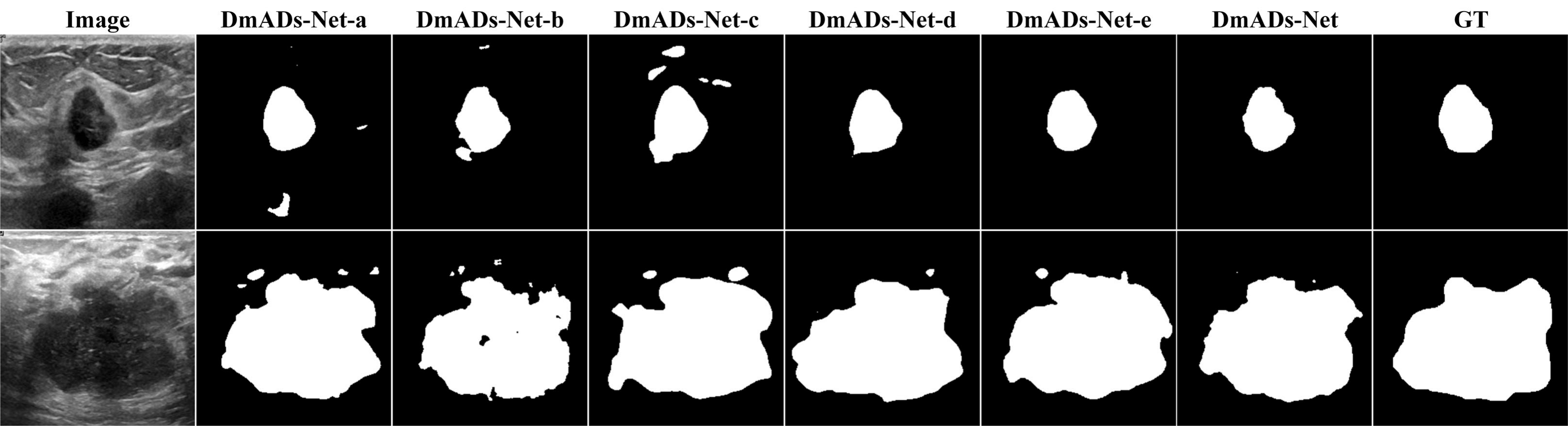}
	\caption{Results of the BUSI dataset on ablation experiment.}
	\label{FIG:26}
\end{figure*}

As in Fig.~\ref{FIG:26} and Table 9, we show ablation experiments performed on the BUSI dataset. From the experimental results, the results obtained by each ablation experiment are not as good as the original network. In particular, the results obtained with DmADs-Net-c contain a large amount of non-target information and are more rounded for lesions, missing information on edge features. The ablation experiments performed on the BUSI dataset similarly demonstrate the effectiveness of the modules in the network.

\begin{table}
	\renewcommand\arraystretch{1.5}
	\caption{Results of the BUSI dataset on ablation experiment}
	\setlength{\tabcolsep}{5.7pt}
	\begin{center}
			\begin{tabular}{lcccc}
					\hline
					\multicolumn{1}{c}{\textbf{Method}} & \textbf{Dice} & \textbf{IoU} & \textbf{Precision} & \textbf{Recall} \\
					\hline
					\textbf{DmADs-Net-a} & 69.41 & 53.16 & 67.54 & 67.66 \\
					\textbf{DmADs-Net-b} & 68.36 & 51.93 & 66.01 & 67.02 \\
					\textbf{DmADs-Net-c} & 70.04 & 53.89 & 73.51 & 66.91 \\
					\textbf{DmADs-Net-d} & 71.20 & 55.28 & 70.79 & 67.88 \\
					\textbf{DmADs-Net-e} & 70.32 & 54.23 & 66.22 & 69.12 \\
					\textbf{DmADs-Net} & 71.35 & 55.46 & 71.14 & 68.46 \\
					\hline
				\end{tabular}
			\label{tab:table9}
		\end{center}
\end{table}

\subsubsection{GlaS}

\begin{figure*}
	\centering
	\includegraphics[scale=0.63]{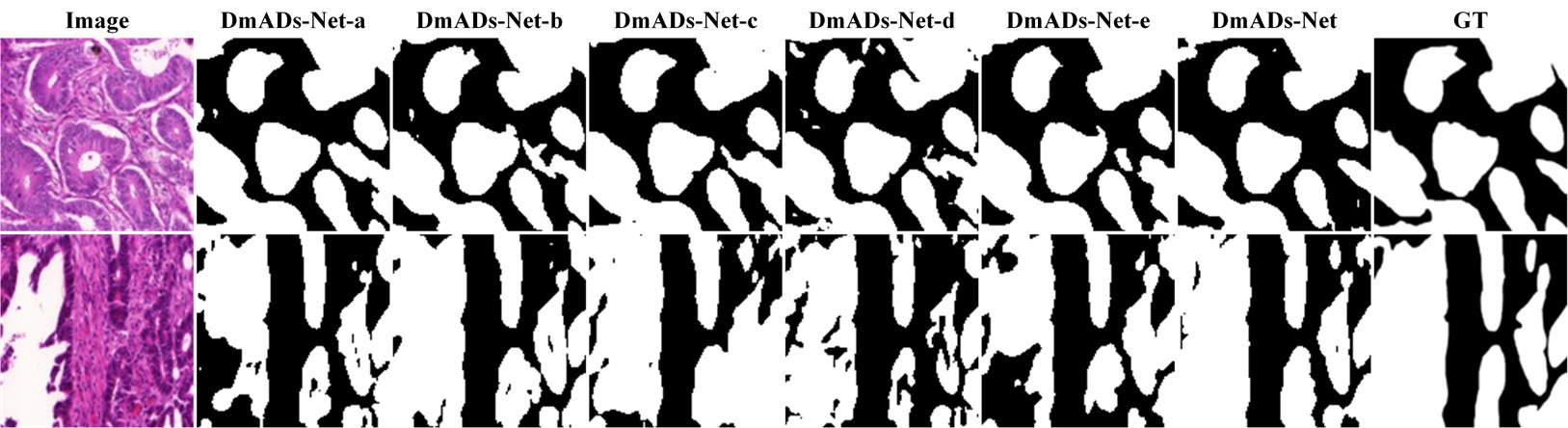}
	\caption{Results of the GlaS dataset on ablation experiment.}
	\label{FIG:27}
\end{figure*}

As in Fig.~\ref{FIG:27}, we modify the sample size of the GlaS dataset to 128$\times$128 and use it for the ablation experiments. Because the lesion targets in the first sample are similar to their surroundings, each ablation experiment exhibits varying degrees of misidentification. For the second sample, it is still the original network that gives the best results. This is indirect evidence of the role played by the three modules and the depth supervision mechanism in the network.

\begin{table}
	\renewcommand\arraystretch{1.5}
	\caption{Results of the GlaS dataset on ablation experiment}
	\setlength{\tabcolsep}{5.7pt}
	\begin{center}
		\begin{tabular}{lcccc}
			\hline
			\multicolumn{1}{c}{\textbf{Method}} & \textbf{Dice} & \textbf{IoU} & \textbf{Precision} & \textbf{Recall} \\
			\hline
			\textbf{DmADs-Net-a} & 88.02 & 79.64 & 89.93 & 87.63 \\
			\textbf{DmADs-Net-b} & 87.30 & 78.60 & 87.28 & 89.06 \\
			\textbf{DmADs-Net-c} & 87.85 & 79.90 & 87.48 & 90.36 \\
			\textbf{DmADs-Net-d} & 86.16 & 76.98 & 85.41 & 88.60 \\
			\textbf{DmADs-Net-e} & 86.73 & 77.32 & 86.94 & 87.49 \\
			\textbf{DmADs-Net} & 88.21 & 79.97 & 90.60 & 87.13 \\
			\hline
		\end{tabular}
		\label{tab:table10}
	\end{center}
\end{table}

As in Table 10, the metric results achieved by the ablation experiments on the GlaS dataset are demonstrated. Combined with the results shown in the comparison plots, performance degradation occurs for all five ablation models, with the most significant degradation occurring for the model with the removal of the depth supervision mechanism.

In the ablation experiments, we conduct experiments on five datasets for fiv ablation models. The ablation experiments demonstrate the contribution of MSCFA to the extraction of weak feature information from the network, the enhancement of the deep supervision mechanism to the network training process, and likewise the contribution of FRFB and LFA to the network performance. Nevertheless, DmADs-Net is still not optimal for some samples, so we still need to explore its optimal performance in the future.

\section{Limitations and future work}

Although DmADs-Net shows good performance on five datasets, we must acknowledge that its universality has not reached the ideal level. Further experiments on other datasets do not demonstrate significant advantages, revealing potential limitations or biases of the model. To enhance the credibility of the paper, future work will delve into these limitations, refine the analysis of the model's performance under different conditions, and identify its performance bottlenecks.

DmADs-Net's performance across various datasets indicates room for improvement in the network's robustness. We plan to conduct in-depth optimization of the network architecture, focusing on improving the capture of global information to enhance overall performance. The integration of the Transformer and diffusion model\cite{55} concepts into DmADs-Net is planned to address current deficiencies.

Considering that most current medical image segmentation datasets are manually created and feature lesions with clear identifiability, real-world complex and emergent medical situations cannot guarantee that lesion targets will be unaffected by environmental factors. Although the datasets used in this paper include samples with environmental interference, their quantity is still insufficient to effectively simulate real situations. In the future, we will validate DmADs-Net's performance in actual medical diagnostics and plan to collect samples affected by environmental interference for systematic model evaluation.

Future research will also be conducted on a broader range of medical image datasets, including specific types such as polyps, retinal blood vessels, and others, as well as in areas like change detection, remote sensing image segmentation, and semantic segmentation. The aim is to expand the applicability of DmADs-Net and adapt it to new application challenges. Through these specific research directions, we aim to comprehensively enhance the performance and applicability of DmADs-Net.

\section{Conclusion}

This paper presents DmADs-Net, aiming to enhance the attention to weak features in medical image segmentation tasks. The Multi-scale Convolutional Feature Attention Block enhances the network's recognition of weak features and improves the intensity of texture and detail information. At the bottleneck layer, the Local Feature Attention Block enhances the capture of high-level semantic information through block processing, thereby improving the overall network performance. The Feature Refinement and Fusion Block, combined with the Edge Spatial Attention strategy, strengthens the fusion of features across different semantic levels. Additionally, a deep supervision mechanism optimizes the training process by calculating the companion loss of intermediate layer outputs, further enhancing the network's learning efficiency. The performance improvement of DmADs-Net in medical image segmentation is expected to have a significant impact on clinical diagnosis and treatment planning. By more accurately identifying and segmenting lesion areas, DmADs-Net can help physicians better understand the condition, leading to more precise diagnoses and more effective treatment plans. For instance, in the early diagnosis of cancer, accurate image segmentation helps in timely tumor detection, increasing the chances of cure. Experimental results prove the efficacy of DmADs-Net and validate the effectiveness of its modules and the rationality of the network structure through ablation studies. In the future, we plan to further explore the field of medical imaging based on DmADs-Net to meet the demands of more medical tasks and advance the development of medical image analysis technology.

%

%


%
%
\bibliographystyle{spmpsci}      
\bibliography{ref.bib}

\begin{figure}[h]
	\includegraphics[width=0.12\textwidth]{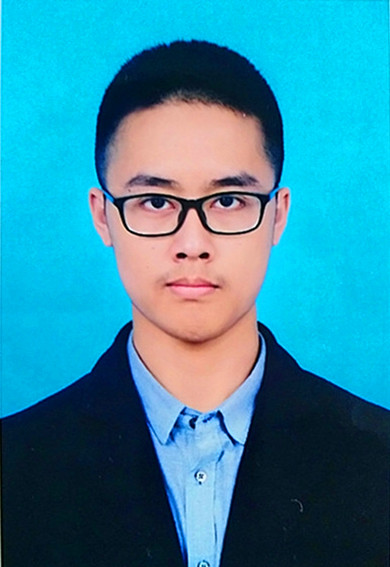}\\
	Zhaojin Fu received his bachelor's degree from the School of Computer Science and Technology, Shandong Technology and Business University, Yantai, China in 2021. Currently studying for a master's degree in the School of Information and Electronic Engineering, Shandong Technology and Business University, Yantai, Shandong. His research interests include computer graphics, computer vision, and image processing.
\end{figure}

\begin{figure}[h]
	\includegraphics[width=0.12\textwidth]{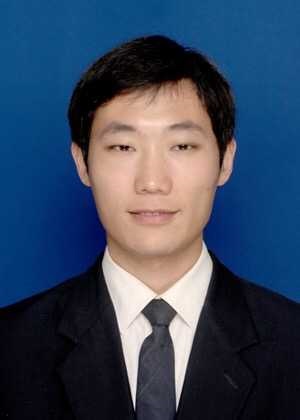}\\
	Zheng Chen received the B.S. and M.S. degrees from Shandong Agricultural University and Shandong Normal University, respectively. He received the Ph.D. degree in Signal and Information Processing from Dalian University of Technology, Dalian, China, in 2022. He is currently a lecturer at Shandong Technology and Business University. His research interests include computer vision, hand pose estimation and hand shape recovery.
\end{figure}

\begin{figure}[h]
	\includegraphics[width=0.12\textwidth]{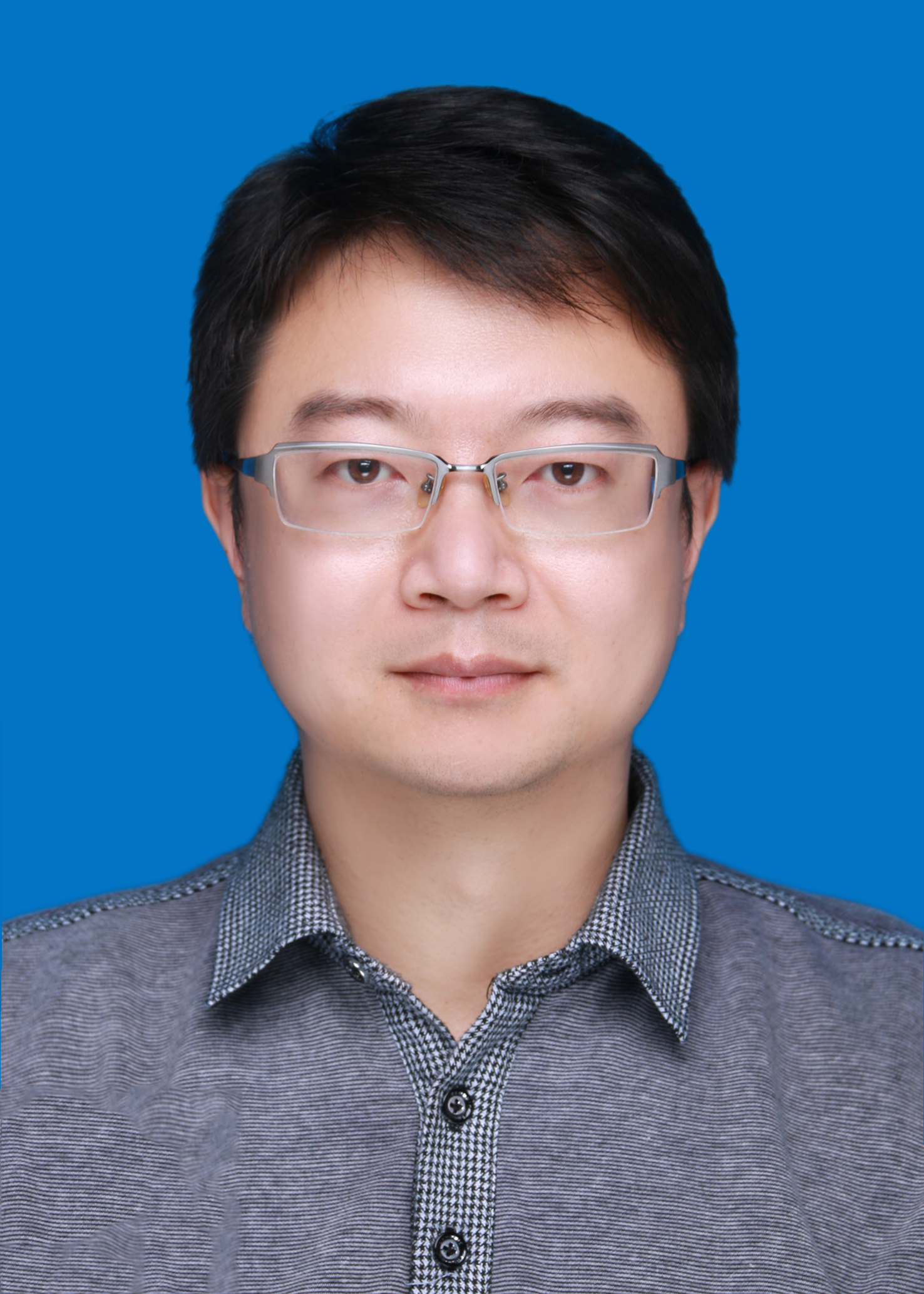}\\
	Jinjiang Li received the B.S. and M.S. degrees in computer science from Taiyuan University of Technology, Taiyuan, China, in 2001 and 2004, respectively, the Ph.D. degree in computer science from Shandong University, Jinan, China, in 2010. From 2004 to 2006, he was an assistant research fellow at the institute of computer science and technology of Peking University, Beijing, China. From 2012 to 2014, he was a Post-Doctoral Fellow at Tsinghua University, Beijing, China. He is currently a Professor at the school of computer science and technology, Shandong Technology and Business University. His research interests include image processing, computer graphics, computer vision and machine learning.
\end{figure}

\begin{figure}[h]
	\includegraphics[width=0.12\textwidth]{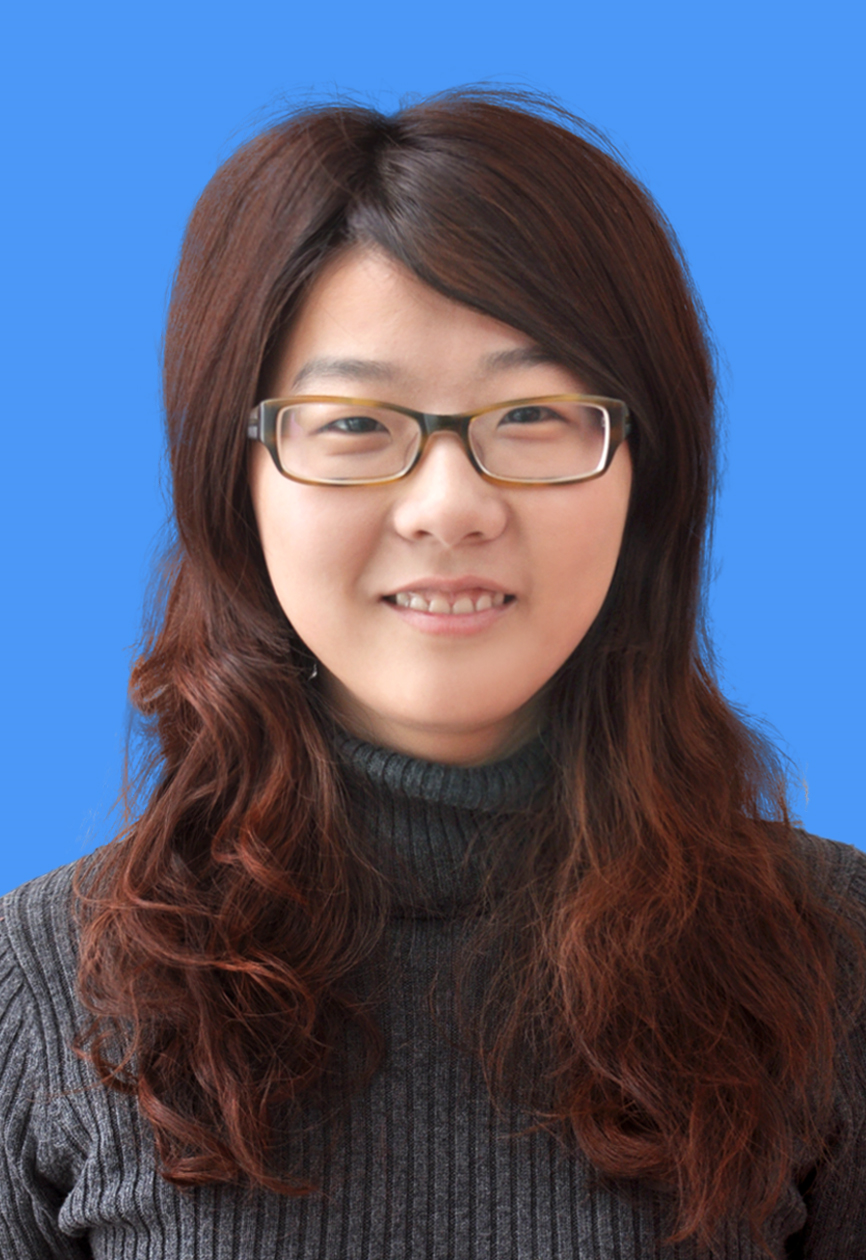}\\
	Lu Ren received the Ph.D. degree in Signal and Information Processing from Dalian University of Technology, Dalian, China. She is currently a lecturer at Shandong Technology and Business University. Her current research interests include sentiment analysis and text mining. 
\end{figure}

\end{sloppypar}
\end{document}